\documentclass[lettersize,journal]{IEEEtran}
\usepackage{amsmath,amsfonts}
\usepackage{algorithmic}
\usepackage{algorithm}
\usepackage{array}

\usepackage{textcomp}
\usepackage{tabularx}
\usepackage{stfloats}
\usepackage{url}
\usepackage{verbatim}
\usepackage{graphicx}
\usepackage{cite}

\usepackage{subcaption} 

\usepackage{makecell}

\usepackage{multirow}
\usepackage[table]{xcolor} 

\definecolor{darkgreen}{RGB}{0, 128, 0}

\definecolor{darkred}{rgb}{255, 0.0, 0.0}

\usepackage{xcolor}
\newcommand{\positive}{\textcolor{darkred}{positive}}
\newcommand{\negative}{\textcolor{darkgreen}{negative}}

\usepackage{amssymb}      
\usepackage{pifont}       

\usepackage{tabularx}     

\usepackage{subcaption}

\usepackage{multirow}
\usepackage[table,xcdraw]{xcolor}
\usepackage{booktabs} 

\usepackage{amsmath} 

\definecolor{declinecolor}{RGB}{204, 0, 0}
\definecolor{increasecolor}{RGB}{0, 153, 0}

\newcommand{\decrease}[1]{\textcolor{declinecolor}{\small$\downarrow$#1}}

\newcommand{\increase}[1]{\textcolor{increasecolor}{\small$\uparrow$#1}}

\begin{document}

\title{FourierCompress: Layer-Aware Spectral Activation Compression for Efficient and Accurate Collaborative LLM Inference}

\author{
    Jian Ma, 
    Xinchen Lyu, 
    Jun Jiang, 
    Longhao Zou,
    Chenshan Ren,
    Qimei Cui, 
    and Xiaofeng Tao
\thanks{

This work was supported in part by Mobile Information Networks National Science and Technology Major Project (Grant No. 2025ZD1303100), and in part by the National Natural Science Foundation of China under Grant 62371059.

Corresponding authors: Xinchen Lyu; Jun Jiang.

        J. Ma, X. Lyu, Q. Cui and X. Tao are with Beijing University of Posts and Telecommunications, China, and also with Pengcheng Laboratory, China.
        J. Jiang and L. Zou is with Pengcheng Laboratory, China.
        C. Ren is with Minzu University of China, China.
    }
}

\markboth{Journal of \LaTeX\ Class Files,~Vol.~14, No.~8, August~2021}%
{Shell \MakeLowercase{\textit{et al.}}: A Sample Article Using IEEEtran.cls for IEEE Journals}

\maketitle

\begin{abstract}
Collaborative large language model (LLM) inference enables real-time, privacy-preserving AI services on resource-constrained edge devices by partitioning computational workloads between client devices and edge servers. However, this paradigm is severely hindered by communication bottlenecks caused by the transmission of high-dimensional intermediate activations, exacerbated by the autoregressive decoding structure of LLMs, where bandwidth consumption scales linearly with output length. Existing activation compression methods struggle to simultaneously achieve high compression ratios, low reconstruction error, and computational efficiency. This paper proposes FourierCompress, a novel, layer-aware activation compression framework that exploits the frequency-domain sparsity of LLM activations. We rigorously demonstrate that activations from the first Transformer layer exhibit strong smoothness and energy concentration in the low-frequency domain, making them highly amenable to near-lossless compression via the Fast Fourier Transform (FFT). FourierCompress transforms activations into the frequency domain, retains only a compact block of low-frequency coefficients, and reconstructs the signal at the server using conjugate symmetry, enabling seamless hardware acceleration on DSPs and FPGAs. Extensive experiments on Llama 3 and Qwen2.5 models across 10 commonsense reasoning datasets demonstrate that FourierCompress preserves performance remarkably close to the uncompressed baseline, outperforming Top-$k$, QR, and SVD. FourierCompress bridges the gap between communication efficiency (an average $7.6\times$ reduction in activation size), near-lossless inference (less than $0.3\%$ average accuracy loss), and significantly faster compression (achieving over $32\times$ reduction in compression time compared to Top-$k$ via hardware acceleration) for edge-device LLM inference.
\end{abstract}

\begin{IEEEkeywords}
LLM, 6G networks, Collaborative LLM Inference, FFT.
\end{IEEEkeywords}

\section{Introduction}

The integration of artificial intelligence (AI) and communication systems stands as one of the six key  scenarios for sixth-generation (6G) wireless networks \cite{wp5d2022future}. A central vision of 6G is to deliver ubiquitous intelligent connectivity, enabling real-time, intelligent services across a vast ecosystem of resource-constrained devices expanding from smartphones and wearables to Internet of Things (IoT) sensors \cite{guo2021enabling,nguyen20216g}. At the forefront of AI advancements are Large Language Models (LLMs), which have demonstrated remarkable capabilities in natural language understanding and generation. However, the deployment on mobile devices faces fundamental limitations due to the immense computational, memory, and energy requirements inherent to billion-parameter LLMs \cite{yu2024edge,zhang2024edgeshard}.

While on-device inference preserves data privacy and minimizes latency, executing state-of-the-art LLMs entirely on mobile platforms remains impractical \cite{ray2024llmedge, li2024collm}. Although techniques such as model distillation \cite{yang2024survey}, quantization \cite{zhu2024survey} , and KV-Cache optimization \cite{liu2024minicache} have been proposed to create lightweight LLMs, the resource demands after model lightweighting still substantially exceed the capabilities of typical mobile devices. This necessitates a distributed inference paradigm (also known as collaborative/split LLM inference) \cite{mudvari2024splitllm}, where the LLM is partitioned into the device portions and edge portions with only intermediate activations transmitted via the wireless channel.

\begin{figure*}
    \centering
    \includegraphics[width=0.75\linewidth]{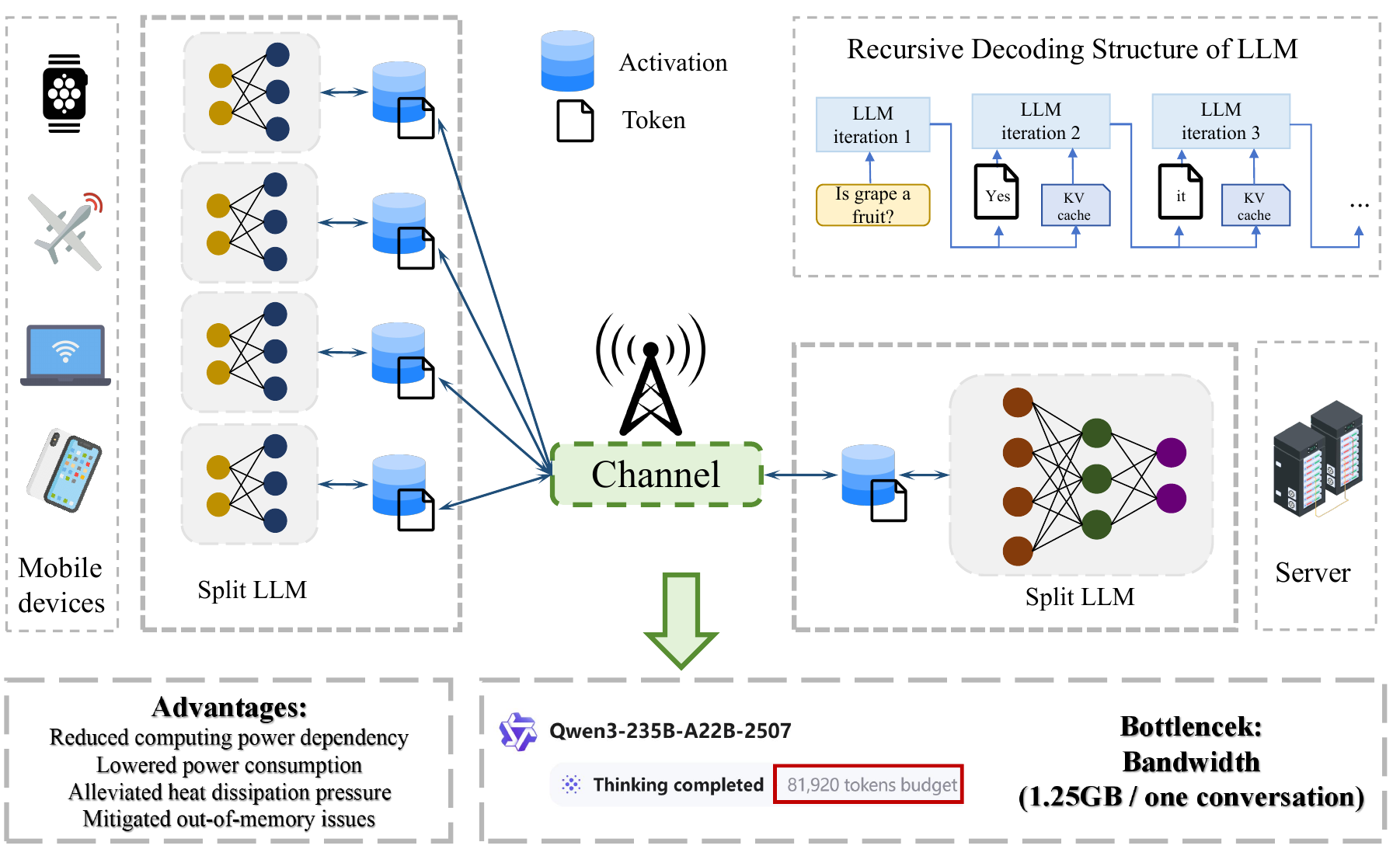}
    \caption{The bandwidth bottleneck of collaborative LLM inference due to recursive decoding structure.}
    \label{fig:figure2}
\end{figure*}

Collaborative LLM inference has garnered significant research interest for enabling real-time, privacy-preserving intelligent, and energy-efficient LLM services on mobile devices \cite{chen2025llm, he2025large,cao2024multimodal,long20246g}. However, as shown in Figure~\ref{fig:figure2}, \textit{the critical challenge of collaborative LLM inference lies in bandwidth consumption of activation transmission arising from the recursive decoding structure of LLMs.} Unlike conventional machine learning models that process inputs in a single forward pass, LLMs generate text token-by-token in an autoregressive manner \cite{vaswani2017attention,achiam2023gpt}. Each new token requires processing the entire history of previously generated tokens, creating a cumulative communication burden that scales linearly with output length. Taking Qwen3-235B as an example, each in-depth thinking conversation consumes approximately 81,920 tokens and requires an activation size of around 1.25 GB. 

Although existing research has explored model partitioning strategies \cite{ong2024efficient}, resource allocation mechanisms \cite{liu2024resource,qian2024user,haider2025llm}, and system optimization techniques \cite{long20246g,he2025large} for collaborative LLM inference, the fundamental bandwidth bottleneck of activation transmission remains inadequately addressed. \textit{This paper aims to address the activation bandwidth bottleneck (essential for both collaborative LLM inference and fine-tuning) by designing an computational efficient compression technique that achieves substantial bandwidth reduction while preserving inference accuracy even under aggressive compression ratios.} 

\subsection{Technical Challenges of Activation Compression}

Traditional compression techniques, such as quantization \cite{lin2024awq,shen2024agile}, sparsification (e.g., Top-$k$) \cite{topk}, and low-rank approximation (e.g., SVD) \cite{fwsvd, asvd, svd-llm}, have been primarily developed for static model weights rather than the transient, data-dependent activations in collaborative LLM inference. This mismatch arises because LLM activations are dynamic, data-driven signals with rich spatial and spectral structure, unlike the relatively stable statistical distributions of model weights \cite{lin2024duquant,an2025systematic}. Consequently, directly applying weight-centric compression to activations results in semantic distortion, where critical information for downstream reasoning is lost despite high compression ratios. Two critical challenges must be addressed to enable efficient and accurate collaborative LLM inference:

\textit{(1) Layer-Specific Compressibility: Where to Split?} A critical yet underexplored question is: which layer serves as the optimal split point for minimizing communication while preserving inference accuracy? Existing approaches typically propose dynamic splitting strategies based on system resources (e.g., ALS \cite{chen2025adaptive}, Splitwise \cite{patel2024splitwise}, and EdgeShard \cite{zhang2024edgeshard}). However, they overlook the intrinsic mathematical properties of activations across different layers. As we demonstrate empirically in Sec. III, activations from different layers exhibit vastly different compressibility. Selecting an appropriate LLM split point is the prerequisite of designing an efficient compression technique.

\textit{(2) Activation-Aware Compression: How to Compress?} Even with an optimal split point, the choice of compression method is crucial. Which compression method can achieve high ratios while preserving the semantic integrity of activations to maintain inference accuracy? Current approaches primarily target model weights rather than activations, and  treat activations as unstructured tensors, leading to suboptimal performance. This creates a critical gap: a lack of activation-specific, layer-aware compression methods for collaborative LLM inference.

\subsection{Contributions and Paper Organization}

We propose FourierCompress, a novel activation compression framework via Fast Fourier Transform (FFT) \cite{FFT} to exploit the spectral sparsity of early-layer LLM activations. We find that activations in the first Transformer layer exhibit strong smoothness and rapid spectral decay with most energy concentrated. We find that in low-frequency components, thus perfectly suitable for compression. By transforming activations into the frequency domain and retaining only the dominant low-frequency coefficients, we achieve high compression ratios with near-lossless reconstruction. FourierCompress addresses both technical challenges:

\textit{(1) Where to Split?} We establish, for the first time, that the first Transformer layer is the optimal split point for activation compression. We provide empirical evidence that early-layer activations are uniquely compressible due to their local attention patterns and structured spatial correlations, whereas deeper layers develop complex, high-entropy representations that resist aggressive compression.

\textit{(2) How to Compress?} We propose a frequency-domain compression framework: transform activations via FFT, retain low-frequency coefficients, and reconstruct using conjugate symmetry. We show that FourierCompress achieves a tighter reconstruction error than its counterparts at the same compression ratio.

From extensive experiments on Llama 3 and Qwen 2.5 across 10 datasets, FourierCompress demonstrate state-of-the-art performance in accuracy, compression ratio, and compression speed.
\begin{itemize}
\item \textit{High Compression Efficiency with Minimal Accuracy Loss.}
By splitting at the first layer, FourierCompress achieves an average 7.6$\times$ reduction in activation size across 10 commonsense reasoning tasks, with less than 0.3\% average accuracy drop, effectively solving the recursive bandwidth problem in autoregressive generation. 
\item \textit{Near-Lossless Inference.} FourierCompress consistently outperforms Top-$k$, QR, SVD, maintaining performance remarkably close to the uncompressed baseline, even surpassing it in some cases.
\item \textit{Hardware-Accelerated Speed.} The use of FFT enables seamless integration with existing hardware accelerators. This reduces compression time by up to 32$\times$ compared to Top-$k$, making it ideal for edge deployment.
\end{itemize}

The rest of this paper is structured as follows. Section II reviews related work in model partitioning and activation compression. Section III provides a layer-specific analysis of activation compressibility to motivate our approach, then details the proposed FourierCompress framework and its hardware integration. Section IV presents extensive experimental results and ablation studies that validate the performance of our method. Finally, Section V concludes the paper.

\section{Related Work}
\label{relatedWork}

This section reviews the recent efforts for collaborative LLM inference, which can be broadly categorized into two complementary directions: (1) where to split the model across edge and server, and (2) how to compress the intermediate activations to reduce communication overhead. As summarized in Table I, existing work typically addresses the two questions separately, and none leverages the signal structure of LLM activations for efficient compression.

\subsection{Model Partitioning: Where to Split?}

\begin{table*}[]
\centering
\caption{Summary of Existing Model Partitioning and Activation Compression Studies.}
\begin{tabularx}{\linewidth}{lll X}
\Xhline{1.2pt} 
\multicolumn{1}{c}{Problem} & \multicolumn{1}{c}{Literature} & \multicolumn{1}{c}{Model Type} & \multicolumn{1}{c}{Details} \\ \hline
& FcaNet \cite{qin2021fcanet} & Small Model & Partitions feature maps by channel to enrich channel attention. \\
& ALS \cite{chen2025adaptive}& LLM & Dynamically splits an LLM by layer between a device and an edge server. \\
& Splitwise \cite{patel2024splitwise}& LLM & Separates prompt computation and token generation onto specialized hardware. \\
& EdgeShard \cite{zhang2024edgeshard}& LLM & Partitions an LLM layer-wise across multiple heterogeneous edge devices. \\
\multirow{-5}{*}{Model Partitioning} & FFSplit \cite{liu2024ffsplit}& LLM & Splits the Feed-Forward Network (FFN) based on neuron output norms. \\ \hline
& COBLA \cite{li2018constrained}& Small Model & Uses constrained optimization to find the optimal low-rank approximation for layers. \\
& Dynamic Pruning \cite{liu2018frequency}& Small Model & Prunes unimportant coefficients in the frequency domain for CNN compression. \\
& SpinQuant \cite{liu2024spinquant}& LLM & Reduces activation outliers using rotation matrices for better low-bit quantization. \\
& Split fine-tuning \cite{topk}& LLM & Compresses activations using Top-$k$ sparsification during distributed training. \\
& AWQ \cite{lin2024awq}& LLM & Protects important weights by scaling them before quantization. \\
& Agile-Quant \cite{shen2024agile}& LLM & Prunes tokens to reduce activation outliers before applying quantization. \\
& Atom \cite{zhao2024atom}& LLM & Reorders activation channels to isolate outliers for mixed-precision quantization. \\
& FWSVD \cite{fwsvd}& LLM & Applies a weighted low-rank factorization based on parameter importance. \\
& Asvd \cite{asvd}& LLM & Transforms weights based on activations for improved low-rank decomposition. \\
\multirow{-10}{*}{Activation Compression} & Svd-llm \cite{svd-llm}& LLM & Uses an activation-based data whitening technique to guide weight decomposition. \\ \hline
\rowcolor[HTML]{D7F9F6} 
\multicolumn{2}{c}{\cellcolor[HTML]{D7F9F6}\textbf{Ours (FourierCompress)}} & \textbf{LLM} & \textbf{Our method addresses both where to split and how to compress by leveraging FFT for high  activation compression ratios and near-lossless reconstruction.} \\ 
\Xhline{1.2pt} 
\end{tabularx}
\end{table*}

Some research has focused on determining the optimal split point in a layered LLM to balance computational load, memory usage, and end-to-end latency. These methods rely on system-level metrics such as network bandwidth, device compute capability, and power constraints: (1) ALS [11] dynamically selects the split layer based on real-time network conditions and node resources; (2) Splitwise [12] uses a phase-splitting strategy to optimize for latency and throughput across multiple inference stages; (3) EdgeShard [13] adaptively partitions the model across heterogeneous edge devices, considering bandwidth and model size; (4) FFSplit [14] proposes splitting within feedforward networks to improve the accuracy-efficiency trade-off.

While these approaches provide valuable system-level optimizations, they overlook the intrinsic mathematical properties of LLM activations. As will be shown in Section III, activation compressibility varies drastically across layers. Splitting at a deep layer can make compression ineffective and degrade accuracy. Our work fills this gap by introducing a signal-aware splitting strategy, establishing for the first time that the first Transformer layer is optimal due to its unique spectral sparsity and spatial smoothness.

\subsection{Activation Compression: How to Compress?}

\textit{(1) Quantization.} Quantization reduces the bit-width of activation values to lower transmission costs. However, quantization alone often provides limited compression ratios without significant accuracy loss, especially in the presence of activation outliers. Many state-of-the-art quantization methods are primarily designed for model weights or require complex calibration. For example, AWQ (Activation-aware Weight Quantization) \cite{lin2024awq} protects salient weights from quantization errors by observing activation magnitudes but is fundamentally a weight-only method. Other techniques, like Agile-Quant \cite{shen2024agile}, address activations by pruning tokens that cause large outliers before applying quantization. While effective, these methods either do not directly compress activations for transmission or rely on removing information (pruning), which can be suboptimal compared to transforming the signal to a more compressible basis.

\textit{(2) Low-Rank Approximation.}
Low-rank approximation methods aim to compress LLM parameter matrices by identifying and preserving the most significant components via Singular Value Decomposition (SVD). However, these techniques are predominantly designed for static model weights, not for the dynamic, data-dependent activations transmitted during collaborative inference. For example, prominent methods like FWSVD \cite{fwsvd}, SVD-LLM \cite{svd-llm}, and ASVD \cite{asvd} all leverage activation characteristics to improve the compression of weight matrices. FWSVD applies a weighted factorization based on parameter importance, SVD-LLM uses an activation-based whitening technique to guide weight decomposition, and ASVD transforms weights based on activation distributions.

\textit{(3) Sparsification.}
Sparsification methods reduce communication by transmitting only a small subset of the most significant activation values. The most prominent technique is Top-$k$ sparsification, which is applied to activations in frameworks likeSplit fine-tuning \cite{topk}. This method retains the $K$ activation values with the largest absolute magnitudes while setting the rest to zero.

\textit{(4) Frequency-Domain Methods.}
Few studies have explored frequency-domain analysis for model compression, and these have been confined to smaller, non-LLM architectures. A notable example is Dynamic Pruning \cite{liu2018frequency}, which prunes unimportant frequency coefficients of \textit{weights} in Convolutional Neural Networks (CNNs) to reduce model size.

Distinctively different from existing work, FourierCompress is the first activation-specific, layer-aware compression approach that jointly considers the split point selection and model compression. By identifying the first Transformer layer as the split point (with energy concentrated in the low-frequency domain), we show that FFT is superior in terms of accuracy, compression ratio, and compression speed. FourierCompress establishes a new paradigm for efficient, nearly-lossless collaborative inference framework with aggressive compression ratio for empowering edge-device LLM services.

\section{Layer-aware Spectral Activation Compression: Motivation and Framework Design}

This section presents the motivation and technical framework of our proposed approach. We first establish the critical insight regarding layer-specific activation compressibility through empirical analysis, and present the FourierCompress framework for efficient collaborative inference.

\begin{figure*}[htbp]
    \centering

    \begin{subfigure}{\textwidth} 
        \centering 
        \includegraphics[width=0.8\linewidth]{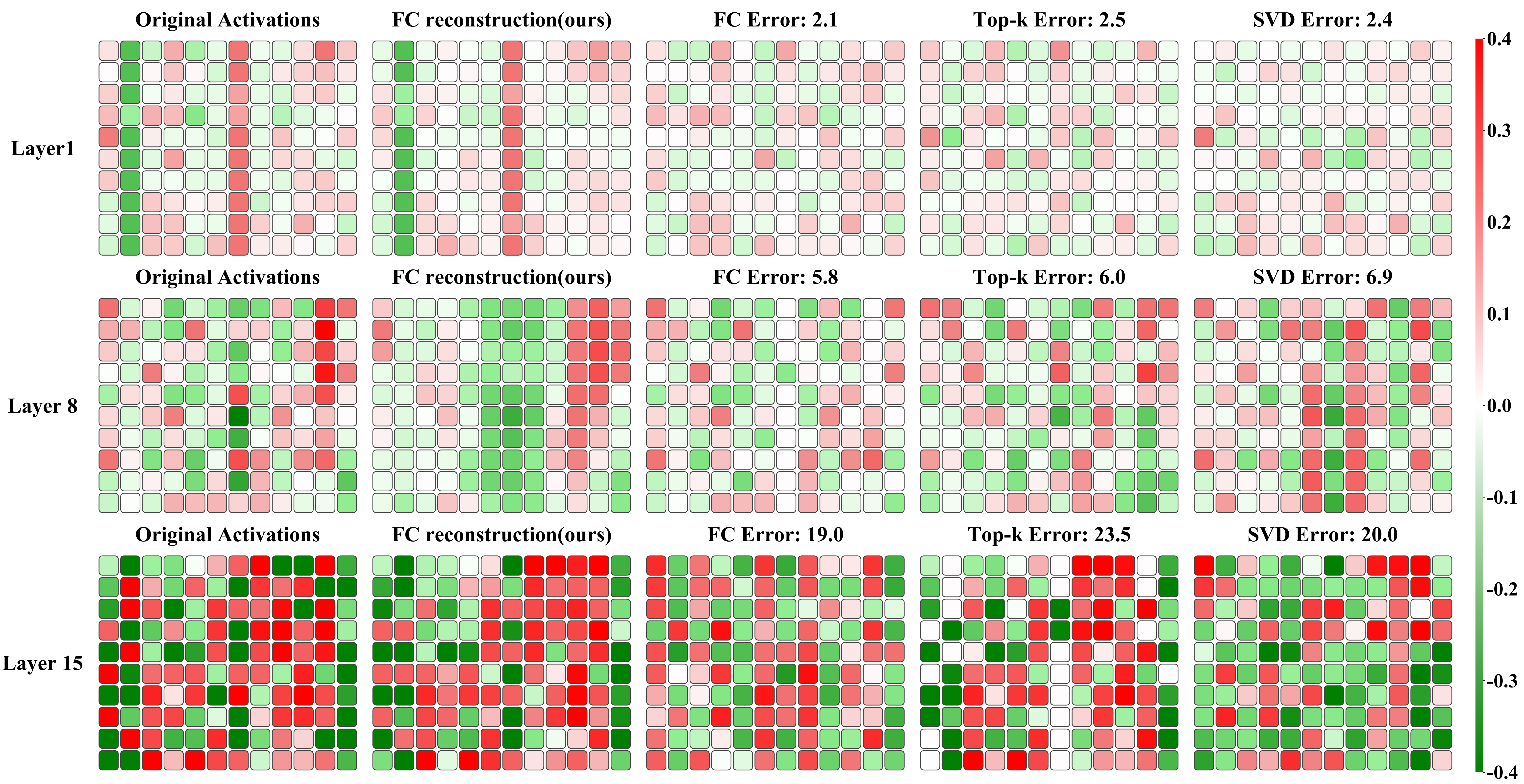}
        \caption{Rows (from top to bottom) show results for different split layers. From left to right is the original activation, the activation reconstructed by FourierCompress, and the errors generated by different compression methods. Coordinate values are color coded (\positive, \negative). As the split layer increases, activation becomes less smooth (more chaotic) and errors also increase.}
        \label{fig:activations_a} 
    \end{subfigure}

    \vspace{1em} 
    
    \begin{subfigure}{\textwidth}
        \centering
        \includegraphics[width=0.8\linewidth]{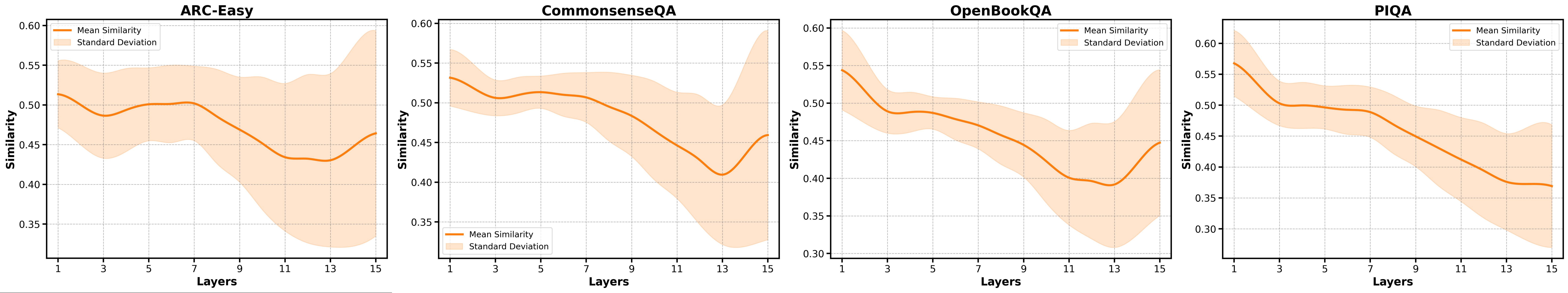}
        \caption{Activation similarity across different datasets(PiQA, ARC-Easy, CommonsenseQA, and OpenBookQA) with increasing layers.}
        \label{fig:activations_b} 
    \end{subfigure}

    \vspace{1em} 

    \begin{subfigure}{\textwidth}
        \centering
        \includegraphics[width=0.65\linewidth]{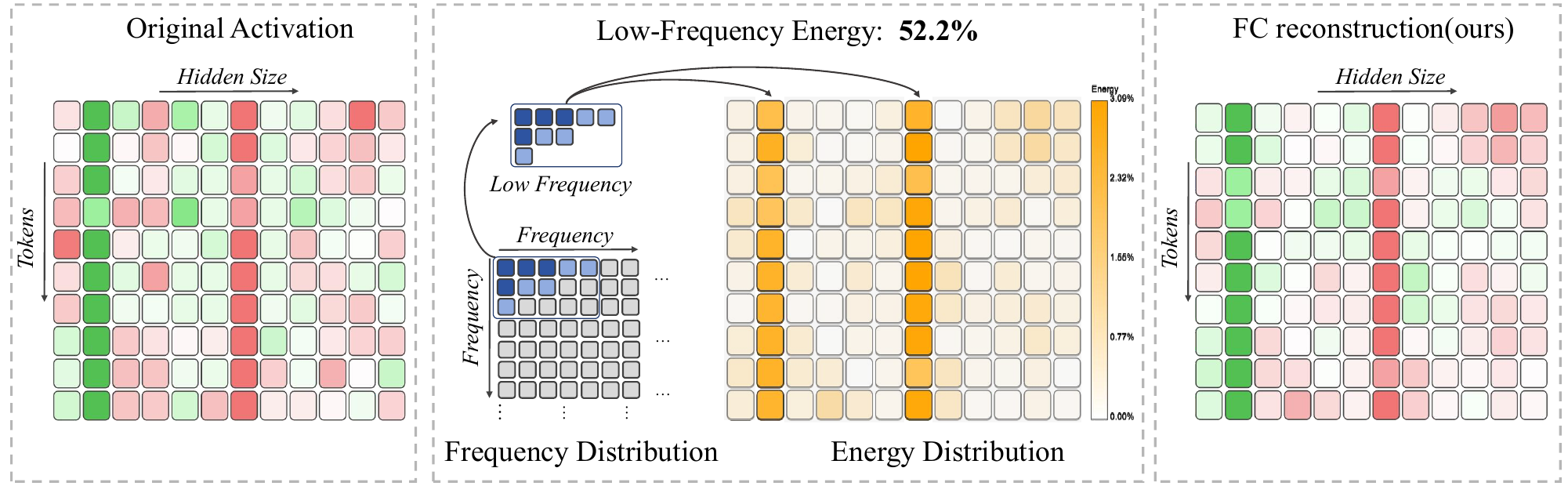}
        \caption{Certain components of the activation, specifically the low-frequency information, exhibit a higher energy distribution, allowing the entire activation to be reconstructed from them.}
        \label{fig:activations_c} 
    \end{subfigure}

    \caption{Activations across different LLM layers exhibit significant differences. Activations at Layer 1 show high similarity in information and energy distribution across different tokens, observed with Llama 3-1B. This property enables FourierCompress to achieve significantly higher compression ratios.}
    \label{fig:activations} 
\end{figure*}

\subsection{Motivation: Layer-Aware Compressibility Analysis}

We start by presenting our key insights regarding the compressibility characteristics of LLM activations, which form the foundation of FourierCompress. We demonstrate through empirical evidence and analysis why early-layer activations exhibit unique properties that make them ideal candidates for spectral compression. This section also provides an overview of the workflow of the FourierCompress framework.

\textit{(1) Analysis of Activation Patterns.} Our investigation begins with an observation that LLM activations exhibit dramatically different compressibility characteristics across network layers \cite{wang2025exploring,li2024adaptive,DBLP:journals/corr/abs-2503-06518}, necessitating a layer-aware compression strategy. Figure~\ref{fig:activations}(a) provides compelling visual evidence of this phenomenon through a comparative analysis of activation patterns and reconstruction errors at different layer depths. As shown in the top row and left column of Figure~\ref{fig:activations}(a), activations from the first Transformer layers display smooth, structured spatial patterns characterized by gradual value transitions and consistent vertical patterns across tokens. These vertical structures indicate that the same neurons activate consistently across different input tokens, reflecting a shared feature extraction mechanism that operates across diverse inputs. In contrast, the bottom row reveals that deeper layer activations exhibit chaotic, high-frequency noise with abrupt value changes and inconsistent activation patterns across tokens.

The smoothness of early-layer activations stems from fundamental properties of the Transformer architecture, particularly the phenomenon of shared feature amplification in early layers. As reported in \cite{wang2025exploring}, when an LLM processes different input tokens, the neuron activation patterns in its first few layers (especially those around the first layer) all show higher similarity.  This strongly indicates that the early layers activate a common set of parameters to perform a foundational analysis of the language, identifying its shared, underlying properties.

In summary, \textit{We find that early-layer activations are uniquely compressible due to their local attention patterns and structured spatial correlations, whereas deeper layers develop complex, high-entropy representations that resist aggressive compression.} The compression error (achieved by FourierCompress and existing  Top-$k$/SVD benchmarks) in right columns of Figure~\ref{fig:activations}(a) validates that reconstruction errors in early layers remain confined to low-magnitude regions, suggesting that the essential semantic information is preserved despite compression. Conversely, deep layers exhibit widespread, high-magnitude errors (intense red patches) that disrupt critical semantic information, even when applying the same compression ratio. This asymmetry demonstrates that a uniform compression approach across all layers is not practical and establishes the necessity for layer-aware compression strategies. Our empirical evidence also demonstrates that the first Transformer layer serves as the optimal split point for activation compression, as its activations maintain maximum structural redundancy while preserving downstream reasoning capabilities.

\textit{(2) Quantitative Analysis of Activation Similarity.} To quantitatively validate our observation, Figure~\ref{fig:activations}(b) presents activation similarity measurements across four diverse datasets (PiQA, ARC-Easy, CommonsenseQA, and OpenBookQA) as a function of layer depth. The results reveal a consistent pattern across all datasets: activation similarity remains high in early layers (peaking around Layers 1-3) but declines sharply as we progress through the network. This trend indicates that early layers function as general, task-agnostic feature extractors, maintaining consistent activation patterns across different inputs and tasks. The high similarity values (approaching 0.8 in some cases) confirm that these layers concentrate information along a few principal directions with significant structural redundancy. As we move to deeper layers, the similarity metrics decrease substantially (often below 0.4), reflecting a transition to complex, highly contextualized representations that are unique to each token's role within the sequence.

Figure~\ref{fig:activations}(b) reveals two critical insights: (1) Early layers exhibit strong cross-dataset consistency in activation patterns, confirming their role as general feature extractors; (2) Deeper layers develop increasingly task-specific representations with lower inherent redundancy, making them inherently less compressible. This transition from general to specific representations explains why early layers are uniquely suitable for high-ratio compression while preserving inference accuracy.

\subsection{Rationale of First-Layer Spectral Activation Compression}
Activations from early Transformer layers, particularly the first layer, exhibit strong spectral concentration in the low-frequency domain. As illustrated in Figure~\ref{fig:activations}(c), the 2D Fourier spectrum of the activation tensor from Layer 1 of Llama 3-1B reveals that the majority of its energy is tightly clustered around the origin (i.e., low-frequency components). This concentration implies that the essential structural and semantic information of the activation can be accurately captured using only a small subset of low-frequency coefficients.

This phenomenon is a direct consequence of the smooth spatial patterns observed in early-layer activations in Figure~\ref{fig:activations}(a), which reflect consistent neuron responses across input tokens. As a result,  by retaining only the dominant low-frequency block submatrix in the top-left corner of the 2D spectrum), spectral compression achieves high compression ratios while preserving the core signal structure. The reconstruction via inverse FFT of this truncated spectrum yields activations that closely approximate the original, as confirmed by the low reconstruction errors in Figure~\ref{fig:activations}(a). This spectral sparsity provides a principled, signal-theoretic justification for applying FFT-based compression specifically at the first Transformer layer, establishing both the layer-awareness and near-lossless fidelity that distinguish FourierCompress from generic compression baselines.

\subsection{Proposed FourierCompress Framework}

\begin{figure*}
    \centering
    \includegraphics[width=0.75\linewidth]{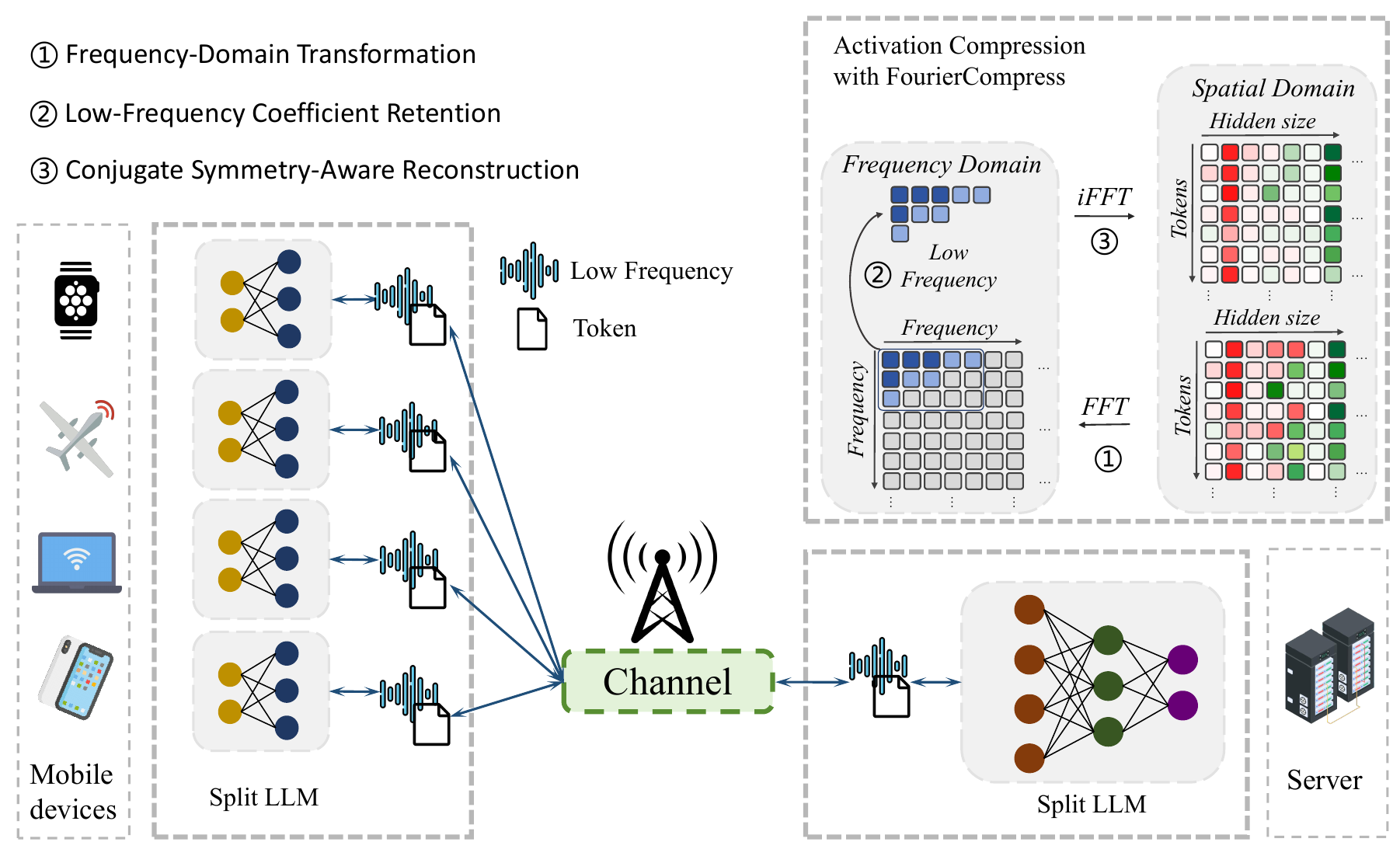}
    \caption{The FourierCompress framework for efficient LLM inference, where the activation is compressed via Fourier transform at the device before transmitting to the edge server.}
    \label{fig:FourierCompressframework}
\end{figure*}  

As shown in Figure~\ref{fig:FourierCompressframework}, the core of our method is to treat the entire activation tensor from the first layer, a matrix whose dimensions correspond to the sequence length ($S$) and the hidden size ($D$), as a single 2D signal. This approach allows us to simultaneously capture correlations both across neurons (the hidden dimension) and between tokens (the sequence dimension). By transforming this 2D signal into the frequency domain, we can isolate and preserve the structurally significant low-frequency components that represent the shared, smooth patterns. The framework consists of three sequential stages: (1) a 2D frequency-domain transformation, (2) low-frequency coefficient retention for compression, and (3) a conjugate symmetry-aware reconstruction. More details are as follows:

\textit{(1) Frequency-Domain Transformation.} For a given input, the activation output of the first Transformer layer is a real-valued matrix $\mathbf{A} \in \mathbb{R}^{S \times D}$, where $S$ is the sequence length and $D$ is the model's hidden size\cite{vaswani2017attention}. We interpret this matrix as a 2D signal, where one axis represents the token sequence and the other represents the neuron activations. The first step is to apply a 2D Fast Fourier Transform (FFT)\cite{FFT} to this entire matrix. This converts the spatio-temporal signal $\mathbf{A}$ into its 2D frequency-domain representation, a matrix of complex numbers $\mathcal{A} \in \mathbb{C}^{S \times D}$, computed as:
\[
\mathcal{A}[u, v] = \sum_{s=0}^{S-1} \sum_{d=0}^{D-1} A[s, d] \cdot e^{-j2\pi(\frac{us}{S} + \frac{vd}{D})}
\]
where $(u, v)$ are the frequency coordinates. This transformation decomposes the activation patterns into constituent frequencies along both the sequence and hidden dimensions. The low-frequency components (where $u$ and $v$ are small), located near the origin of the 2D spectrum, represent the smooth, dominant patterns that are consistent across both tokens and neurons, effectively capturing the shared structure.

\textit{(2) Low-Frequency Coefficient Retention.} With the activations transformed into a 2D frequency spectrum, compression is achieved by retaining a rectangular block of the most informative, low-frequency coefficients. We select cutoff points $K_S$ and $K_D$ (where $K_S \ll S$ and $K_D \ll D$) based on the target compression ratio. We then preserve only the top-left $K_S \times K_D$ block of coefficients from the spectrum $\mathcal{A}$. This truncation strategy is principled: it keeps the coefficients corresponding to low frequencies in both dimensions, which encode the essential shared structure, while discarding the high-frequency coefficients that represent less critical, token-specific and neuron-specific variations. By transmitting only this smaller $K_S \times K_D$ matrix of complex numbers, the data volume is substantially reduced.

\textit{(3) Conjugate Symmetry-Aware Reconstruction.} On the receiving end, the original $S \times D$ activation matrix is reconstructed from the received $K_S \times K_D$ block of coefficients. This process relies on the fact that the 2D Fourier transform of a real-valued matrix exhibits conjugate symmetry ($\mathcal{A}[u, v] = \mathcal{A}^*[S-u, D-v]$), making the high-frequency components inherently redundant. In practice, the reconstruction is performed efficiently by creating a new $S \times D$ matrix, placing the received $K_S \times K_D$ block in the top-left corner, and padding the rest of the matrix with zeros. This zero-padded matrix, $\mathcal{A}_{\text{padded}}$, effectively serves as a low-pass filtered version of the original spectrum. Finally, the server applies the 2D Inverse Fast Fourier Transform (IFFT) to $\mathcal{A}_{\text{padded}}$ to obtain the reconstructed activation matrix $\mathbf{A}' \in \mathbb{R}^{S \times D}$:
\[
A'[s, d] = \frac{1}{SD} \sum_{u=0}^{S-1} \sum_{v=0}^{D-1} \mathcal{A}_{\text{padded}}[u, v] \cdot e^{j2\pi(\frac{us}{S} + \frac{vd}{D})}
\]
This method provides a highly efficient, metadata-free reconstruction, as the position of the coefficients is implicitly known.

\subsection{Hardware Integration and Performance Analysis}

FourierCompress is designed not only for high compression fidelity but also for practical deployment on resource-constrained edge platforms. Its core operation FFT benefits from decades of hardware optimization and is natively supported across a wide range of edge accelerators, including DSPs and FPGAs. This section details our hardware-aware implementation strategy and analyzes its impact on system-level performance.

\textit{(1) Hardware Acceleration.} The computational complexity of FourierCompress is dominated by the 2D FFT and its inverse, both of which scale as $\mathcal{O}(SD \log(SD))$, where $ S $ is the sequence length and $ D $ is the hidden dimension. This is substantially more efficient than the $\mathcal{O}(S D^2) $ or higher complexity of low-rank methods like SVD or QR decomposition. More importantly, FFT is highly amenable to parallelization and fixed-function hardware acceleration, and particularly suitable for edge deployment. We implemented and evaluated FourierCompress in two representative edge hardware environments:

\begin{itemize}
    \item \textit{GPU/DSP Acceleration:} On NVIDIA Jetson platforms, we leveraged cuFFT, a highly optimized CUDA library, to perform real-time FFT/IFFT operations. This yields significant speedups over general-purpose CPU execution.
    \item \textit{FPGA Implementation:} On the Alinx AXU15EGB FPGA platform, we deployed a custom FFT core using a pipelined architecture that exploits dedicated DSP slices. This design achieves over 10$\times$ throughput improvement compared to software baselines by maximizing data reuse and minimizing memory access latency.
\end{itemize}

\textit{(2) Performance Evaluation Metrics.} The hardware-aware design of FourierCompress translates into dramatic reductions in compression latency. As will shortly shown in Section IV, our hardware-accelerated implementation reduces activation compression time by up to 32$\times$ compared to Top-$k$ sparsification and by over two orders of magnitude compared to SVD-based approaches. Crucially, this acceleration incurs only a 0.3\% average overhead in total end-to-end inference latency, making it suitable for real-time edge applications.

This combination of algorithmic efficiency, near-lossless reconstruction, and seamless hardware integration establishes FourierCompress as a practical and scalable solution for mitigating the communication bottleneck in collaborative LLM inference, particularly in emerging 6G edge AI scenarios where bandwidth, latency, and energy are tightly constrained.

\section{Experimental Results}
\label{sec:experiments}

This section presents a comprehensive empirical evaluation of FourierCompress (FC) across four different LLM models, ten popular datasets, and FPGA/Jetson hardware platforms. We assess its performance in terms of inference accuracy, compression ratio, computational overhead, and system-level scalability under realistic collaborative inference scenarios.

\subsection{Experimental Setup}

\begin{figure}
    \centering
    \includegraphics[width=0.95\linewidth]{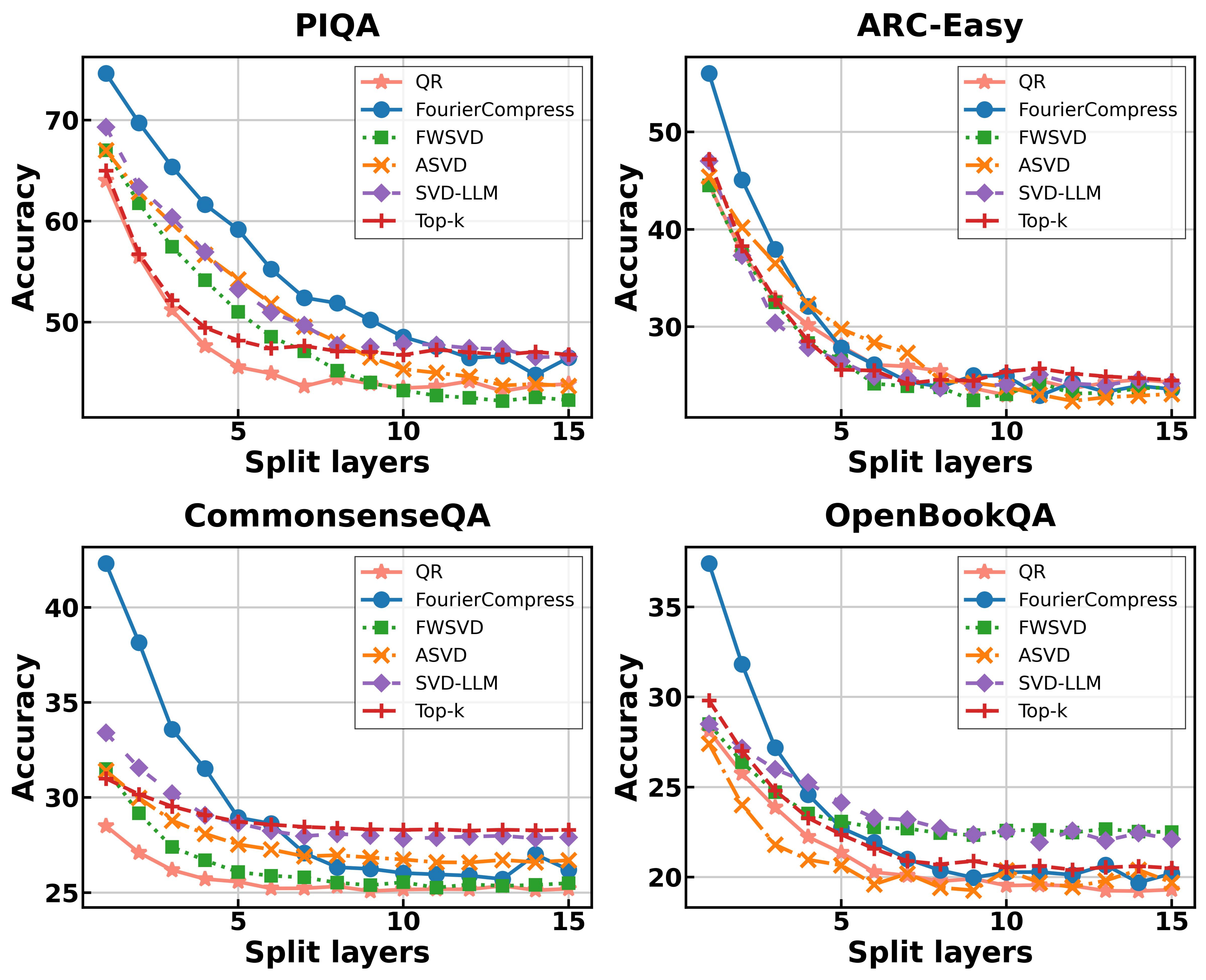}
    \caption{Comparison of Split Layer and Accuracy in Llama 3.}
    \label{fig_layer_acc}
\end{figure}  

\begin{figure}
    \centering
    \includegraphics[width=0.95\linewidth]{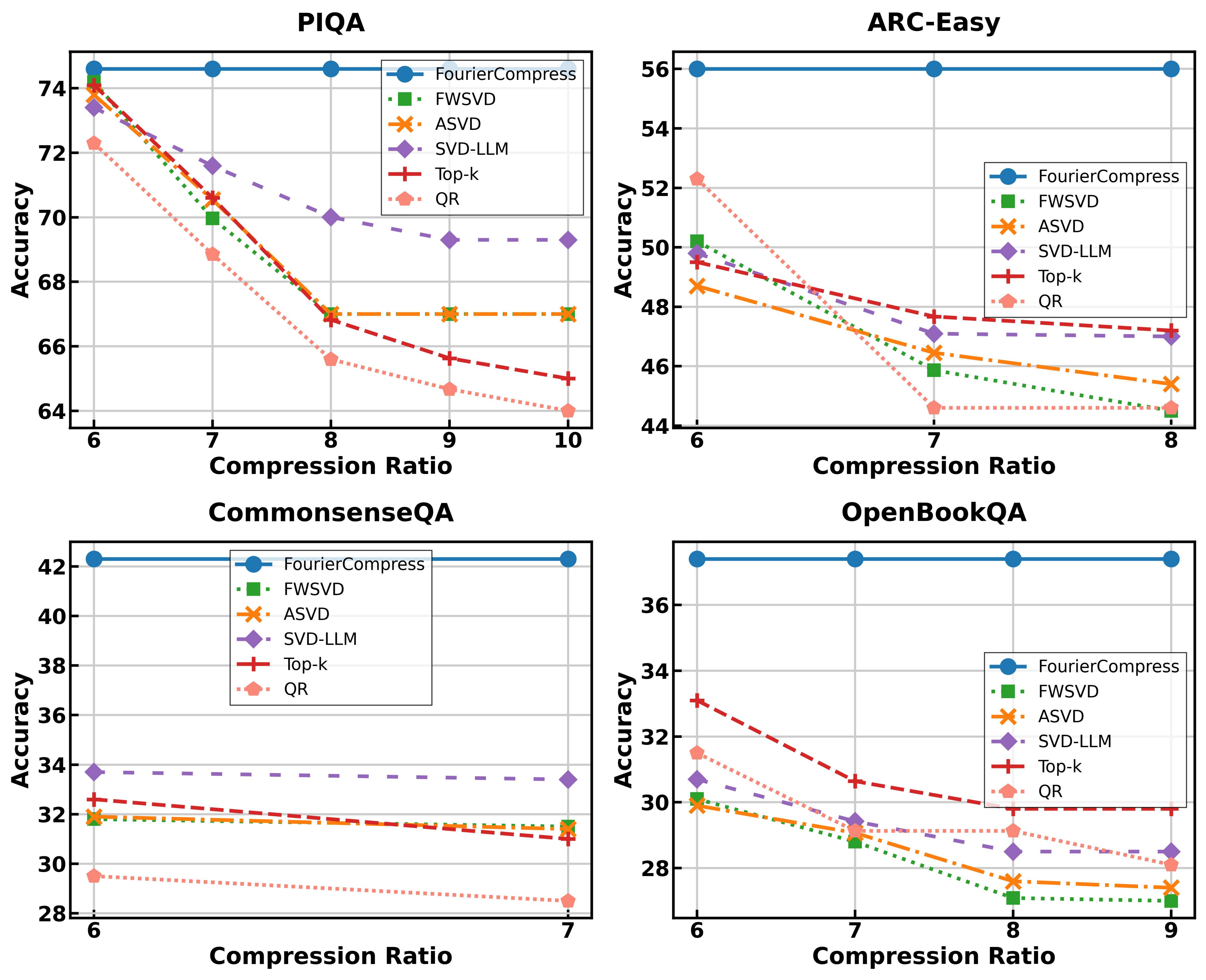}
    \caption{Comparison of Compression Ratio and Accuracy in Llama 3.}
    \label{fig_ratio_acc}
\end{figure}

\textbf{Models and Datasets:}
We conducted experiments using two families of LLMs: Llama 3\cite{llama3modelcard} (specifically, Llama 3-1B and Llama 3-3B) and Qwen2.5\cite{qwen2.5} (Qwen2.5-1.5B and Qwen2.5-3B). To assess the generalization capabilities of the compression methods, we evaluated them on a comprehensive suite of 10 commonsense reasoning datasets: OpenBookQA\cite{OpenBookQA2018} (OA), ARC-Easy\cite{allenai:arc} (A-e), ARC-Challenge\cite{allenai:arc} (A-c), PIQA\cite{Bisk2020} (PA), SIQA\cite{sap2019socialiqa} (SA), WinoGrande\cite{ai2:winogrande} (WG), CommonsenseQA\cite{talmor-etal-2019-commonsenseqa} (CQ), QASC\cite{allenai:qasc} (QC), LogiQA\cite{liu2020logiqa} (LA), and CosmosQA\cite{huang-etal-2019-cosmos} (CA). The primary evaluation metric is accuracy (reported in percentages).

\textbf{Hardware and software:} The experimental environment featured eight NVIDIA RTX 4090 GPUs for general simulation tasks and establishing performance baselines. Jetson-DSP and Zynq-FPGA were used for the accelerated computation of FourierCompress, simulating its performance on edge devices. Core software included Python 3.11, PyTorch 2.7.0 (built with CUDA 11.8).

\begin{table*}[]
\caption{Different datasets have different FourierCompress ratios. Other methods were also evaluated using the same compression ratios as FourierCompress for fair comparison.} 
\centering
\begin{tabular}{cccccccccccc}
\Xhline{1.2pt}
Model                       & Compression Ratio    & \textbf{OA}  & \textbf{A-e} & \textbf{A-c} & \textbf{PA}  & \textbf{SA}  & \textbf{WG}  & \textbf{CQ}  & \textbf{QC}  & \textbf{LA}  & \textbf{CA}  \\ \hline
                            & 10       & 36.5         & 53.7         & 37.4         & \cellcolor[HTML]{D7F9F6}74.6 & 35.4         & 54.3         & 40.9         & 20.7         & 30.6         & 23.5         \\
                            & 9        & \cellcolor[HTML]{D7F9F6}37.4 & 55.2         & 38.1         & \cellcolor[HTML]{EBFCFB}74.6 & 38.7         & 55.2         & 41.1         & 21.2         & 32.1         & 25.3         \\
                            & 8        & \cellcolor[HTML]{EBFCFB}37.4 & \cellcolor[HTML]{D7F9F6}56.0 & \cellcolor[HTML]{D7F9F6}39.5 & \cellcolor[HTML]{EBFCFB}74.6 & 41.8         & 56.1         & 41.2         & 21.2         & \cellcolor[HTML]{D7F9F6}34.6 & 26.5         \\
                            & 7        & \cellcolor[HTML]{EBFCFB}37.4 & \cellcolor[HTML]{EBFCFB}56.2 & \cellcolor[HTML]{EBFCFB}39.7 & \cellcolor[HTML]{EBFCFB}74.6 & 42.5         & 56.3         & \cellcolor[HTML]{D7F9F6}42.3 & \cellcolor[HTML]{D7F9F6}22.3 & \cellcolor[HTML]{EBFCFB}34.1 & \cellcolor[HTML]{D7F9F6}27.1 \\
                            & 6        & \cellcolor[HTML]{EBFCFB}37.4 & \cellcolor[HTML]{EBFCFB}56.4 & \cellcolor[HTML]{EBFCFB}39.8 & \cellcolor[HTML]{EBFCFB}74.6 & \cellcolor[HTML]{D7F9F6}44.2 & \cellcolor[HTML]{D7F9F6}57.8 & \cellcolor[HTML]{EBFCFB}42.3 & \cellcolor[HTML]{EBFCFB}22.4 & \cellcolor[HTML]{EBFCFB}33.3 & \cellcolor[HTML]{EBFCFB}27.3 \\
\multirow{-6}{*}{Llama 3-1B}   & Baseline & \cellcolor[HTML]{EBFCFB}37.4 & \cellcolor[HTML]{EBFCFB}56.5 & \cellcolor[HTML]{EBFCFB}39.8 & \cellcolor[HTML]{EBFCFB}74.6 & \cellcolor[HTML]{EBFCFB}44.0 & \cellcolor[HTML]{EBFCFB}57.9 & \cellcolor[HTML]{EBFCFB}42.5 & \cellcolor[HTML]{EBFCFB}22.4 & \cellcolor[HTML]{EBFCFB}33.3 & \cellcolor[HTML]{EBFCFB}27.3 \\ \hline
                            & 10       & 38.5         & 58.7         & 39.4         & \cellcolor[HTML]{D7F9F6}76.5 & 40.1         & 56.5         & 42.3         & 22.2         & 35.2         & 26.8         \\
                            & 9        & \cellcolor[HTML]{D7F9F6}40.8 & 59.3         & 40.3         & \cellcolor[HTML]{EBFCFB}76.4 & 42.9         & 56.9         & 43.4         & 23.6         & 37.4         & 28.4         \\
                            & 8        & \cellcolor[HTML]{EBFCFB}41.3 & \cellcolor[HTML]{D7F9F6}62.1 & \cellcolor[HTML]{D7F9F6}42.1 & \cellcolor[HTML]{EBFCFB}76.4 & 45.8         & 59.7         & 43.1         & 24.2         & \cellcolor[HTML]{D7F9F6}39.5 & 30.5         \\
                            & 7        & \cellcolor[HTML]{EBFCFB}41.2 & \cellcolor[HTML]{EBFCFB}62.5 & \cellcolor[HTML]{EBFCFB}42.5 & \cellcolor[HTML]{EBFCFB}76.4 & \cellcolor[HTML]{D7F9F6}46.9 & 60.4         & \cellcolor[HTML]{D7F9F6}44.3 & \cellcolor[HTML]{D7F9F6}25.8 & \cellcolor[HTML]{EBFCFB}40.1 & \cellcolor[HTML]{D7F9F6}31.4 \\
                            & 6        & \cellcolor[HTML]{EBFCFB}41.6 & \cellcolor[HTML]{EBFCFB}62.5 & \cellcolor[HTML]{EBFCFB}42.8 & \cellcolor[HTML]{EBFCFB}76.4 & \cellcolor[HTML]{EBFCFB}46.6 & \cellcolor[HTML]{D7F9F6}63.5 & \cellcolor[HTML]{EBFCFB}44.3 & \cellcolor[HTML]{EBFCFB}26.0 & \cellcolor[HTML]{EBFCFB}40.5 & \cellcolor[HTML]{EBFCFB}31.4 \\
\multirow{-6}{*}{Llama 3-3B}   & Baseline & \cellcolor[HTML]{EBFCFB}41.6 & \cellcolor[HTML]{EBFCFB}62.5 & \cellcolor[HTML]{EBFCFB}42.8 & \cellcolor[HTML]{EBFCFB}76.4 & \cellcolor[HTML]{EBFCFB}46.6 & \cellcolor[HTML]{EBFCFB}63.5 & \cellcolor[HTML]{EBFCFB}44.5 & \cellcolor[HTML]{EBFCFB}26.0 & \cellcolor[HTML]{EBFCFB}40.7 & \cellcolor[HTML]{EBFCFB}31.4 \\ \hline
                            & 10       & \cellcolor[HTML]{D7F9F6}39.6 & 60.8         & 40.2         & \cellcolor[HTML]{D7F9F6}73.7 & 44.3         & 56.7         & 30.8         & 21.2         & 40.2         & 20.7         \\
                            & 9        & \cellcolor[HTML]{EBFCFB}39.1 & 61.6         & 41.3         & \cellcolor[HTML]{EBFCFB}73.8 & 44.7         & 56.8         & 30.4         & 21.2         & 40.5         & 21.5         \\
                            & 8        & \cellcolor[HTML]{EBFCFB}38.4 & \cellcolor[HTML]{D7F9F6}62.5 & \cellcolor[HTML]{D7F9F6}42.8 & \cellcolor[HTML]{EBFCFB}73.8 & 46.2         & 57.2         & 31.7         & 21.6         & \cellcolor[HTML]{D7F9F6}40.7 & 22.1         \\
                            & 7        & \cellcolor[HTML]{EBFCFB}38.4 & \cellcolor[HTML]{EBFCFB}63.5 & \cellcolor[HTML]{EBFCFB}44.1 & \cellcolor[HTML]{EBFCFB}73.8 & 48.4         & 57.2         & \cellcolor[HTML]{D7F9F6}33.0 & \cellcolor[HTML]{D7F9F6}21.9 & \cellcolor[HTML]{EBFCFB}41.1 & \cellcolor[HTML]{D7F9F6}22.6 \\
                            & 6        & \cellcolor[HTML]{EBFCFB}38.4 & \cellcolor[HTML]{EBFCFB}63.7 & \cellcolor[HTML]{EBFCFB}44.2 & \cellcolor[HTML]{EBFCFB}73.8 & \cellcolor[HTML]{D7F9F6}50.2 & \cellcolor[HTML]{D7F9F6}57.7 & \cellcolor[HTML]{EBFCFB}33.5 & \cellcolor[HTML]{EBFCFB}22.3 & \cellcolor[HTML]{EBFCFB}41.8 & \cellcolor[HTML]{EBFCFB}22.6 \\
\multirow{-6}{*}{Qwen2.5-1.5B} & Baseline & \cellcolor[HTML]{EBFCFB}38.4 & \cellcolor[HTML]{EBFCFB}63.7 & \cellcolor[HTML]{EBFCFB}44.2 & \cellcolor[HTML]{EBFCFB}73.8 & \cellcolor[HTML]{EBFCFB}49.1 & \cellcolor[HTML]{EBFCFB}58.8 & \cellcolor[HTML]{EBFCFB}33.7 & \cellcolor[HTML]{EBFCFB}22.3 & \cellcolor[HTML]{EBFCFB}42.0 & \cellcolor[HTML]{EBFCFB}22.6 \\ \hline
                            & 10       & 38.7         & 62.1         & 42.2         & \cellcolor[HTML]{D7F9F6}75.7 & 43.1         & 59.2         & 32.3         & 23.5         & 35.7         & 23.7         \\
                            & 9        & \cellcolor[HTML]{D7F9F6}40.4 & \cellcolor[HTML]{D7F9F6}64.4 & 42.6         & \cellcolor[HTML]{EBFCFB}75.7 & 45.7         & 60.4         & 32.4         & 23.8         & 36.6         & 25.4         \\
                            & 8        & \cellcolor[HTML]{EBFCFB}40.4 & \cellcolor[HTML]{EBFCFB}64.0 & \cellcolor[HTML]{D7F9F6}43.5 & \cellcolor[HTML]{EBFCFB}75.6 & 48.3         & 62.5         & \cellcolor[HTML]{D7F9F6}34.7 & 25.6         & \cellcolor[HTML]{D7F9F6}37.0 & 26.6         \\
                            & 7        & \cellcolor[HTML]{EBFCFB}40.4 & \cellcolor[HTML]{EBFCFB}64.0 & \cellcolor[HTML]{EBFCFB}43.5 & \cellcolor[HTML]{EBFCFB}75.7 & 48.6         & 62.8         & \cellcolor[HTML]{EBFCFB}34.2 & 26.7         & \cellcolor[HTML]{EBFCFB}35.8 & \cellcolor[HTML]{D7F9F6}27.2 \\
                            & 6        & \cellcolor[HTML]{EBFCFB}40.4 & \cellcolor[HTML]{EBFCFB}64.0 & \cellcolor[HTML]{EBFCFB}43.5 & \cellcolor[HTML]{EBFCFB}75.7 & \cellcolor[HTML]{D7F9F6}49.2 & \cellcolor[HTML]{D7F9F6}64.0 & \cellcolor[HTML]{EBFCFB}34.2 & \cellcolor[HTML]{D7F9F6}27.9 & \cellcolor[HTML]{EBFCFB}35.9 & \cellcolor[HTML]{EBFCFB}27.6 \\
\multirow{-6}{*}{Qwen2.5-3B}       & Baseline & \cellcolor[HTML]{EBFCFB}40.4 & \cellcolor[HTML]{EBFCFB}64.0 & \cellcolor[HTML]{EBFCFB}43.5 & \cellcolor[HTML]{EBFCFB}75.7 & \cellcolor[HTML]{EBFCFB}49.8 & \cellcolor[HTML]{EBFCFB}63.9 & \cellcolor[HTML]{EBFCFB}34.2 & \cellcolor[HTML]{EBFCFB}27.9 & \cellcolor[HTML]{EBFCFB}35.8 & \cellcolor[HTML]{EBFCFB}27.6 \\ \hline
\rowcolor[HTML]{ECF4FF} 
\multicolumn{2}{c}{\cellcolor[HTML]{ECF4FF}\textbf{Avg.  Compression Ratio}} & \textbf{8.7} & \textbf{8.4} & \textbf{8.3} & \textbf{10.3} & \textbf{6}   & \textbf{5.8} & \textbf{7.5} & \textbf{6.8} & \textbf{8.0} & \textbf{6.6} \\ 

\Xhline{1.2pt}
\end{tabular}

\label{tab:compressRatio} 
\end{table*}

\textbf{Compared Methods:}
We compare our proposed FourierCompress framework against a comprehensive suite of representative compression techniques, all applied directly to the intermediate activation tensors for a fair evaluation. The performance upper bound is established by a Baseline, where no compression is applied. We evaluate the following methods:
\begin{enumerate}
    \item \textbf{Top-$k$}: \cite{topk} retains only the $k$ activation values with the largest absolute magnitudes for sparsification.
    \item \textbf{FWSVD} \cite{fwsvd}: A state-of-the-art SVD variant which applies a weighted factorization.
    \item \textbf{ASVD} \cite{asvd}: A state-of-the-art SVD variant which transforms weights based on activation statistics. 
    \item \textbf{SVD-LLM} \cite{svd-llm}: A state-of-the-art SVD variant which uses data whitening to guide decomposition. 
    \item \textbf{QR Decomposition (QR)} \cite{Zhuang2024Medical}: A classical low-rank factorization method included alongside the SVD variants.
    \item \textbf{Baseline} : No compression (upper-bound performance).
\end{enumerate}

\subsection{Accuracy and Compression Performance}

\textit{(1) Validation of Early-Layer Splitting.} Figure~\ref{fig_layer_acc} validates our core design principle: splitting at the first Transformer layer is essential for high-fidelity activation compression.  On Llama 3-1B, we compare the performance of FourierCompress against other methods across four datasets (PIQA, OpenBookQA, CommonsenseQA, and ARC-Easy) at their respective optimal compression ratios (ranging from 7.5$\times$ to 10.3$\times$; more details in Table~\ref{tab:compressRatio}). At the first layer, all methods achieve relatively high accuracy, with FourierCompress being the most accurate. However, as the split layer moves to deeper layers (e.g., layer 5 or 15), the accuracy of all methods drops sharply. For instance, on PIQA, FourierCompress’s accuracy falls from 74.6\% at Layer 1 to 48\% at Layer 15. This sharp degradation confirms that deeper activations lack the structural redundancy required for aggressive compression, reinforcing the necessity of layer-aware splitting.

\textit{(2) Dataset-Adaptive Near-Lossless Compression Ratios.} Table~\ref{tab:compressRatio} presents a fine-grained ablation to identify the maximum compression ratio that incurs less than 0.3\% accuracy loss relative to the uncompressed baseline—our definition of “near-lossless.” Results reveal significant dataset-dependent variability: PIQA supports up to 10.3$\times$ compression due to its high spectral redundancy, whereas WinoGrande is more sensitive, tolerating only 5.8$\times$. \textbf{The average compression ratio across all 10 datasets is 7.6$\times$}, which we adopt as the standard evaluation setting for fair comparison in Table~\ref{tab:acc}.


\textit{(3) Accuracy Comparison at Fixed Compression Ratios.} Table~\ref{tab:acc} presents a comparison of inference accuracy, with each method evaluated at the compression ratios determined in Table~\ref{tab:compressRatio}. \textbf{FourierCompress consistently preserves accuracy to within 0.3\% of the uncompressed baseline across all four models and ten datasets.} In stark contrast, competing methods suffer substantial degradation under the same aggressive compression ratios. For instance, on Llama 3-1B with the CommonsenseQA dataset, FourierCompress achieves 42.3\% accuracy (a negligible 0.2-point drop), whereas SVD-LLM and Top-$k$ plummet to 33.4\% and 31.0\%, respectively—a performance loss of over 9 points. This highlights a fundamental limitation of methods like Top-$k$, which disrupt the spatial coherence of the activation tensor, and SVD-based approaches, which are ill-suited for capturing the complex structures inherent in activations. Remarkably, on Qwen2.5-3B, FourierCompress's average accuracy (46.4\%) slightly exceeds the baseline (46.3\%). We attribute this to a beneficial regularization effect; by design, FourierCompress acts as a low-pass filter, discarding high-frequency components that may represent noise. This implicit denoising can improve the signal-to-noise ratio for subsequent layers, thereby enhancing generalization.

\textit{(4) Robustness Across Compression Ratios.} Figure~\ref{fig_ratio_acc} examines the accuracy–compression trade-off on Llama 3. FourierCompress exhibits graceful degradation. In stark contrast, SVD-based methods collapse rapidly—SVD-LLM’s accuracy drops by more than 8 points at the same ratio—highlighting FourierCompress’s superior ability to retain semantically critical information under aggressive compression. This robustness stems directly from its exploitation of the strong low-frequency energy concentration in early-layer activations (Figure~\ref{fig:activations}(c)), which ensures that discarded coefficients contribute minimally to overall signal fidelity.

\begin{table*}[]

\caption{Comparison of Accuracy at the Same Compression Ratio.}
\centering
\begin{tabular}{llccccccccccc}
\toprule[1.2pt]
\textbf{Model} & \textbf{Method} & \textbf{OA} & \textbf{A-e} & \textbf{A-c} & \textbf{PA} & \textbf{SA} & \textbf{WG} & \textbf{CQ} & \textbf{QC} & \textbf{LA} & \textbf{CA} & \textbf{Avg.} \\
\midrule
& FWSVD \cite{fwsvd} & 27.0 & 44.5 & 29.8 & 67.0 & 41.4 & 52.9 & 31.5 & 19.9 & 32.3 & 24.0 & 37.0 (\decrease{6.6}) \\
& ASVD \cite{asvd}& 27.4 & 45.4 & 31.1 & 67.0 & 41.1 & 53.9 & 31.4 & 20.1 & 32.8 & 23.7 & 37.4 (\decrease{6.2})\\
& SVD-LLM \cite{svd-llm}& 28.5 & 47.0 & 32.0 & 69.3 & 43.9 & 55.6 & 33.4 & 20.9 & 33.2 & 25.1 & 38.9 (\decrease{4.7}) \\
& QR \cite{Zhuang2024Medical}& 28.1 & 44.6 & 33.3 & 64.0 & 37.2 & 49.2 & 28.5 & 18.6 & 31.6 & 22.5 & 35.8 (\decrease{7.8})\\
& Top-$k$ \cite{topk}& 29.8 & 47.2 & 29.8 & 65.0 & 40.6 & 53.0 & 31.0 & 19.3 & 32.1 & 23.3 & 37.1 (\decrease{6.5})\\
& \cellcolor[HTML]{D7F9F6}FC & \cellcolor[HTML]{D7F9F6}\textbf{37.4} & \cellcolor[HTML]{D7F9F6}56.0 & \cellcolor[HTML]{D7F9F6}39.5 & \cellcolor[HTML]{D7F9F6}\textbf{74.6} & \cellcolor[HTML]{D7F9F6}\textbf{44.2} & \cellcolor[HTML]{D7F9F6}57.8 & \cellcolor[HTML]{D7F9F6}42.3 & \cellcolor[HTML]{D7F9F6}22.3 & \cellcolor[HTML]{D7F9F6}\textbf{34.6} & \cellcolor[HTML]{D7F9F6}27.1 & \cellcolor[HTML]{D7F9F6}\textbf{43.6 (\decrease{0.0})} \\
\multirow{-7}{*}{Llama 3-1B} & Baseline & \textbf{37.4} & \textbf{56.5} & \textbf{39.8} & \textbf{74.6} & 44.0 & \textbf{57.9} & \textbf{42.5} & \textbf{22.4} & 33.3 & \textbf{27.3} & \textbf{43.6} \\
\midrule
& FWSVD \cite{fwsvd}& 27.8 & 52.7 & 38.1 & 73.7 & 44.3 & 62.1 & 43.6 & 23.9 & 39.2 & 30.4 & 43.6 (\decrease{4.0})\\
& ASVD \cite{asvd}& 28.1 & 52.8 & 38.1 & 73.6 & 44.3 & 62.4 & 43.9 & 24.1 & 39.2 & 30.4 & 43.7 (\decrease{3.9})\\
& SVD-LLM \cite{svd-llm}& 28.3 & 53.3 & 38.7 & 73.4 & 44.8 & 62.2 & 43.5 & 24.3 & 40.9 & 30.9 & 44.0 (\decrease{3.6})\\
& QR \cite{Zhuang2024Medical}& 31.5 & 53.4 & 37.7 & 74.3 & 41.5 & 59.1 & 41.5 & 23.2 & 40.2 & 31.5 & 43.4 (\decrease{4.2})\\
& Top-$k$ \cite{topk}& 38.6 & 64.2 & 41.1 & 75.3 & 45.0 & 62.8 & 44.1 & 24.0 & 40.7 & 32.3 & 46.8 (\decrease{0.8})\\
& \cellcolor[HTML]{D7F9F6}FC & \cellcolor[HTML]{D7F9F6}40.8 & \cellcolor[HTML]{D7F9F6}62.1 & \cellcolor[HTML]{D7F9F6}42.1 & \cellcolor[HTML]{D7F9F6}\textbf{76.5} & \cellcolor[HTML]{D7F9F6}\textbf{46.9} & \cellcolor[HTML]{D7F9F6}\textbf{63.5} & \cellcolor[HTML]{D7F9F6}44.3 & \cellcolor[HTML]{D7F9F6}25.8 & \cellcolor[HTML]{D7F9F6}39.5 & \cellcolor[HTML]{D7F9F6}\textbf{31.4} & \cellcolor[HTML]{D7F9F6}47.3 \textbf{(\decrease{0.3})}\\
\multirow{-7}{*}{Llama 3-3B} & Baseline & \textbf{41.6} & \textbf{62.5} & \textbf{42.8} & 76.4 & 46.6 & \textbf{63.5} & \textbf{44.5} & \textbf{26.0} & \textbf{40.7} & \textbf{31.4} & \textbf{47.6} \\
\midrule
& FWSVD \cite{fwsvd}& 30.2 & 48.9 & 31.6 & 72.6 & 49.1 & 57.6 & 32.4 & 21.0 & 39.2 & 22.0 & 40.5 (\decrease{4.3})\\
& ASVD \cite{asvd}& 31.1 & 50.2 & 32.2 & 72.6 & 49.6 & 57.1 & 32.2 & 20.1 & 39.9 & 22.4 & 40.7 (\decrease{4.1})\\
& SVD-LLM \cite{svd-llm}& 31.9 & 51.5 & 33.3 & 72.8 & 50.0 & 57.6 & 33.3 & 21.1 & 40.7 & 22.3 & 41.4 (\decrease{3.4})\\
& QR \cite{Zhuang2024Medical}& 31.6 & 47.4 & 41.5 & 70.5 & 46.4 & 55.1 & 31.5 & 19.9 & 37.1 & 22.1 & 40.3 (\decrease{4.5})\\
& Top-$k$ \cite{topk}& 38.0 & 62.2 & 42.5 & 72.8 & 48.1 & 56.9 & 32.0 & 21.3 & 38.5 & 21.9 & 43.4 (\decrease{1.4})\\
& \cellcolor[HTML]{D7F9F6}FC & \cellcolor[HTML]{D7F9F6}\textbf{39.6} & \cellcolor[HTML]{D7F9F6}62.5 & \cellcolor[HTML]{D7F9F6}42.8 & \cellcolor[HTML]{D7F9F6}73.7 & \cellcolor[HTML]{D7F9F6}\textbf{50.2} & \cellcolor[HTML]{D7F9F6}57.7 & \cellcolor[HTML]{D7F9F6}33.0 & \cellcolor[HTML]{D7F9F6}21.9 & \cellcolor[HTML]{D7F9F6}40.7 & \cellcolor[HTML]{D7F9F6}\textbf{22.6} & \cellcolor[HTML]{D7F9F6}44.5 \textbf{(\decrease{0.3})}\\
\multirow{-7}{*}{QWen2.5-1.5B} & Baseline & 38.4 & \textbf{63.7} & \textbf{44.2} & \textbf{73.8} & 49.1 & \textbf{58.8} & \textbf{33.7} & \textbf{22.3} & \textbf{42.0} & \textbf{22.6} & \textbf{44.8} \\
\midrule
& FWSVD \cite{fwsvd}& 29.3 & 50.5 & 36.9 & 73.6 & 48.9 & 62.8 & 33.5 & 26.4 & 33.1 & 26.5 & 42.1 (\decrease{4.2})\\
& ASVD \cite{asvd}& 29.2 & 50.3 & 36.9 & 73.2 & 48.6 & 62.2 & 33.2 & 26.4 & 33.1 & 26.3 & 41.9 (\decrease{4.4})\\
& SVD-LLM \cite{svd-llm}& 29.7 & 51.2 & 37.5 & 74.3 & 49.8 & 63.8 & 34.0 & 26.9 & 33.7 & 26.8 & 42.8 (\decrease{3.5})\\
& QR \cite{Zhuang2024Medical}& 27.5 & 54.9 & 38.5 & 73.3 & 48.2 & 62.7 & 32.9 & 26.9 & 32.6 & 25.9 & 42.3 (\decrease{4.0})\\
& Top-$k$ \cite{topk}& 39.2 & 62.6 & 42.2 & 74.1 & 48.9 & 63.4 & 33.6 & 27.7 & 33.3 & 26.6 & 45.2 (\decrease{1.1})\\
& \cellcolor[HTML]{D7F9F6}FC & \cellcolor[HTML]{D7F9F6}\textbf{40.4} & \cellcolor[HTML]{D7F9F6}\textbf{64.4} & \cellcolor[HTML]{D7F9F6}\textbf{43.5} & \cellcolor[HTML]{D7F9F6}\textbf{75.7} & \cellcolor[HTML]{D7F9F6}49.2 & \cellcolor[HTML]{D7F9F6}\textbf{64.0} & \cellcolor[HTML]{D7F9F6}\textbf{34.7} & \cellcolor[HTML]{D7F9F6}\textbf{27.9} & \cellcolor[HTML]{D7F9F6}\textbf{37.0} & \cellcolor[HTML]{D7F9F6}27.2 & \cellcolor[HTML]{D7F9F6}\textbf{46.4 \increase{0.1}} \\
\multirow{-7}{*}{QWen2.5-3B} & Baseline & \textbf{40.4} & 64.0 & \textbf{43.5} & \textbf{75.7} & \textbf{49.8} & 63.9 & 34.2 & \textbf{27.9} & 35.8 & \textbf{27.6} & 46.3 \\
\bottomrule[1.2pt]
\end{tabular}

\label{tab:acc}
\end{table*}

\subsection{Compression Efficiency and Hardware Acceleration}

\begin{table*}[]
\centering
\caption{Total time for activation compression and decompression (s).}
\begin{tabular}{ccccccccc}
\Xhline{1.2pt} 
LLM          & Hidden Size & FWSVD & ASVD  & SVD-LLM & QR     & Top-$k$ & FC(software) & FC (hardware) \\ \hline
Llama 3-1B    & 2048        & 90.4  & 197.8 & 94.2    & 2354.0 & 22.8  & 6.9          & 1.0           \\
Llama 3-3B    & 3072        & 327.6 & 212.7 & 141.8   & 2076.4 & 30.9  & 6.7          & 0.7           \\
Qwen2.5-1.5B & 1536        & 141.8 & 113.4 & 81.0    & 2027.5 & 15.6  & 6.9          & 0.6           \\
Qwen2.5-3B   & 2048        & 11.5  & 171.5 & 89.6    & 2057.9 & 20.4  & 4.9          & 0.5           \\
Avg.         & —           & 142.8 & 173.9 & 101.7   & 2129.0 & 22.4  & \textbf{6.4} & \textbf{0.7}  \\ \Xhline{1.2pt}
\end{tabular}

\label{tab:table3}
\end{table*}

\begin{figure}
    \centering
    \includegraphics[width=0.75\linewidth]{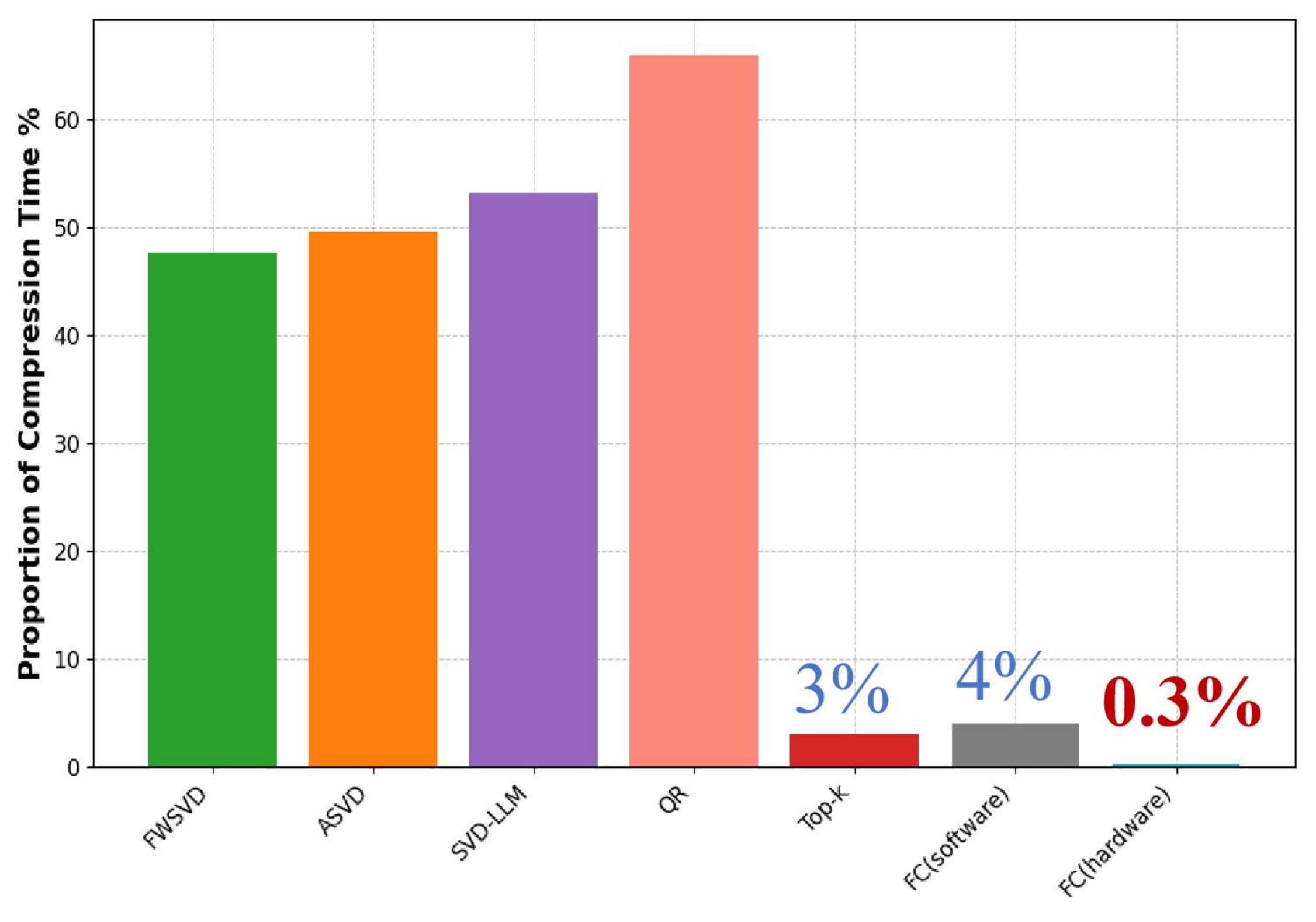}
    \caption{Proportion of compression time in the response time.}
    \label{fig:zhanbi}
\end{figure}

\begin{figure*}[htbp]
    \centering 

    \begin{subfigure}{0.9\textwidth} 
        \centering 
        \includegraphics[width=\linewidth]{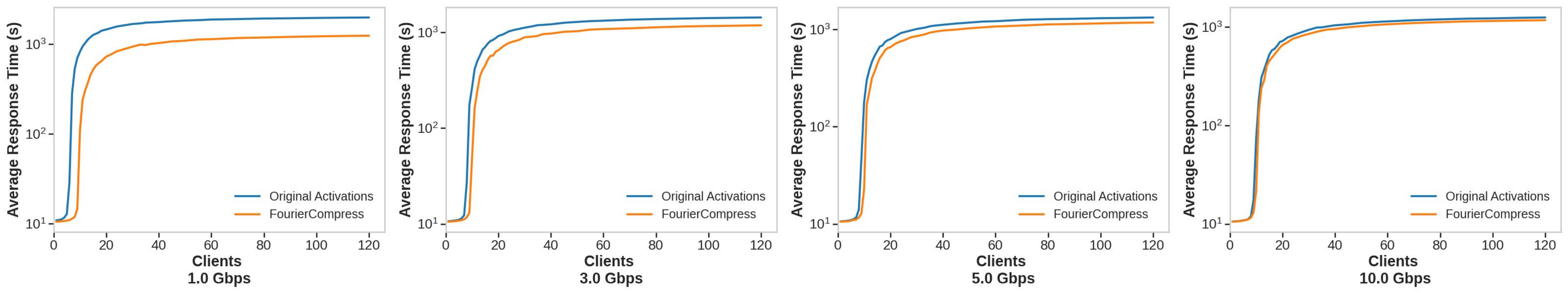}
        \caption{The server has an NVIDIA GeForce RTX 4090. As the number of clients increases, the average response time grows rapidly for both Original Activations and FC. This response time is not reduced by improving network speed.}
        \label{fig:weak_server} 
    \end{subfigure}
    
    \vspace{1em} 

    \begin{subfigure}{0.9\textwidth} 
        \centering 
        \includegraphics[width=\linewidth]{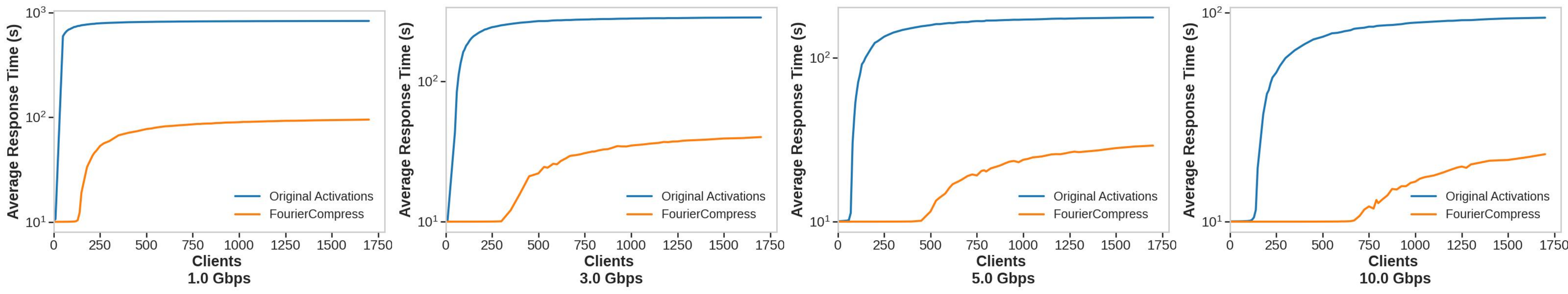}
        \caption{The server has 8 NVIDIA GeForce RTX 4090s. As the number of clients increases, the average response time for Original Activations grows rapidly, while FC reduces the average response time. With higher network speeds, FC can support more clients.}
        \label{fig:strong_server} 
    \end{subfigure}
    
    \caption{Comparison of Average Client Response Time with Different Server Computational Capabilities and Network Speeds (Gbps).}
    \label{fig:4090} 
\end{figure*}

To assess the practical viability of FourierCompress for real-time edge deployment, we evaluate both the absolute compression/decompression latency and its contribution to end-to-end inference time. As shown in Table~\ref{tab:table3}, traditional low-rank methods incur prohibitive computational overhead. For instance, QR decomposition requires an average of 2129.0s, and SVD-LLM takes 101.7 s to compress activations from input across four LLMs (Llama 3-1B/3B and Qwen2.5-1.5B/3B). Even the relatively lightweight Top-$k$ sparsification demands 22.4s on average, which remains impractical for interactive applications.

In contrast, the software implementation of FourierCompress achieves a compression time of only 6.4s—already 3.5$\times$ faster than Top-$k$ and $>$15$\times$ faster than SVD-LLM. This efficiency stems from the near-linear complexity of the 2D FFT (O(SDlog(SD)) ) and its highly parallelizable structure. More importantly, FourierCompress is uniquely suited for hardware acceleration. By leveraging dedicated FFT engines on edge platforms—such as cuFFT on NVIDIA Jetson GPUs or custom pipelined FFT cores on FPGAs—the compression time drops dramatically to an average of 0.7 s. \textbf{FC represents 32$\times$ speedup over Top-$k$} and over two orders of magnitude improvement compared to SVD-based approaches.

Figure~\ref{fig:zhanbi} contextualizes this gain within the full inference pipeline, showing the proportion of compression time relative to total response latency (including client-side processing, transmission, and server-side computation). Low-rank methods dominate the latency budget: QR and SVD-LLM consume 66\% and 53\% of total time, respectively. While Top-$k$ and software-based FourierCompress introduce modest overheads (3\% and 4\%), hardware-accelerated FourierCompress reduces this to just 0.3\%. This near-negligible overhead confirms that FourierCompress is not only accurate and communication-efficient but also computationally lightweight—making it ideal for real-time, resource-constrained edge AI systems.

\subsection{Multi-Client Scalability Under Varying 6G Network Conditions}

To evaluate the practical impact of FourierCompress in real-world collaborative inference scenarios, we simulate a multi-client edge computing environment under different 6G wireless data rates (1 Gbps, 3 Gbps, 5 Gbps, and 10 Gbps), using the Llama 3 model and the PIQA dataset. We analyze system behavior under two distinct resource regimes to identify the dominant performance bottleneck, as illustrated in Figure~\ref{fig:4090}.

\textit{(1) Computation-constrained Regime.} In Figure~\ref{fig:4090}(a), the edge server is equipped with a single NVIDIA RTX 4090 GPU. Here, computational capacity—not communication bandwidth—limits system throughput. As the number of concurrent clients increases beyond approximately 10, the server becomes saturated, causing average response time to rise sharply for both the uncompressed baseline and FourierCompress. Crucially, improving network speed yields negligible latency reduction in this regime, confirming that when computation is the bottleneck, activation compression provides minimal benefit.

\textit{(2) Bandwidth-constrained Regime.} In contrast, Figure~\ref{fig:4090}(b) depicts a bandwidth-constrained scenario where the server is scaled to eight RTX 4090 GPUs, providing ample compute resources. Under this configuration, communication becomes the primary bottleneck. FourierCompress dramatically reduces average client response time across all tested network speeds by shrinking activation payloads by an average of 10.3$\times$. Moreover, it significantly enhances system scalability: at 10 Gbps, the system supports over 1,500 concurrent clients before saturation, compared to only 150 clients with uncompressed activations. This near 10$\times$ increase in client capacity demonstrates that FourierCompress is not just an incremental improvement but a critical enabling technology. It effectively shifts the system bottleneck away from the communication channel and back to the more easily scalable domain of computation. This allows system operators to fully exploit the high-throughput capabilities of future 6G networks, making efficient and large-scale collaborative LLM inference a practical reality.

\section{Conclusion}

FourierCompress effectively addresses the critical trade-off between communication efficiency, inference accuracy, and compression overhead. By strategically exploiting the unique spectral characteristics of early-layer LLM activations, it offers a robust and efficient pathway towards enabling scalable and privacy-preserving collaborative LLM inference on edge devices.

\bibliographystyle{IEEEtran}

\bibliography{references}

\begin{thebibliography}{10}
\providecommand{\url}[1]{#1}
\csname url@samestyle\endcsname
\providecommand{\newblock}{\relax}
\providecommand{\bibinfo}[2]{#2}
\providecommand{\BIBentrySTDinterwordspacing}{\spaceskip=0pt\relax}
\providecommand{\BIBentryALTinterwordstretchfactor}{4}
\providecommand{\BIBentryALTinterwordspacing}{\spaceskip=\fontdimen2\font plus
\BIBentryALTinterwordstretchfactor\fontdimen3\font minus \fontdimen4\font\relax}
\providecommand{\BIBforeignlanguage}[2]{{%
\expandafter\ifx\csname l@#1\endcsname\relax
\typeout{** WARNING: IEEEtran.bst: No hyphenation pattern has been}%
\typeout{** loaded for the language `#1'. Using the pattern for}%
\typeout{** the default language instead.}%
\else
\language=\csname l@#1\endcsname
\fi
#2}}
\providecommand{\BIBdecl}{\relax}
\BIBdecl

\bibitem{wp5d2022future}
I.~{WP5D}, ``Future technology trends of terrestrial international mobile telecommunications systems towards 2030 and beyond,'' \emph{International Telecommunication Union, Report M}, pp. 2516--0, 2022.

\bibitem{guo2021enabling}
F.~Guo, F.~R. Yu, H.~Zhang, X.~Li, H.~Ji, and V.~C. Leung, ``{Enabling} {Massive} {IoT} {Toward} {6G}: A {Comprehensive} {Survey},'' \emph{{IEEE} Internet of Things Journal}, vol.~8, no.~15, pp. 11\,891--11\,915, 2021.

\bibitem{nguyen20216g}
D.~C. Nguyen, M.~Ding, P.~N. Pathirana, A.~Seneviratne, J.~Li, D.~Niyato, O.~Dobre, and H.~V. Poor, ``{6G} {Internet} of {Things}: A {Comprehensive} {Survey},'' \emph{{IEEE} Internet of Things Journal}, vol.~9, no.~1, pp. 359--383, 2021.

\bibitem{yu2024edge}
Z.~Yu, Z.~Wang, Y.~Li, R.~Gao, X.~Zhou, S.~R. Bommu, Y.~Zhao, and Y.~Lin, ``{EDGE-LLM}: {Enabling} {Efficient} {Large} {Language} {Model} {Adaptation} on {Edge} {Devices} via {Unified} {Compression} and {Adaptive} {Layer} {Voting},'' in \emph{Proceedings of the 61st {ACM/IEEE} Design Automation Conference}, 2024, pp. 1--6.

\bibitem{zhang2024edgeshard}
M.~Zhang, X.~Shen, J.~Cao, Z.~Cui, and S.~Jiang, ``{EdgeShard}: {Efficient} {LLM} {Inference} via {Collaborative} {Edge} {Computing},'' \emph{{IEEE} Internet of Things Journal}, 2024.

\bibitem{ray2024llmedge}
P.~P. Ray and M.~P. Pradhan, ``{LLMEdge}: {A} {Novel} {Framework} for {Localized} {LLM} {Inferencing} at {Resource} {Constrained} {Edge},'' in \emph{2024 International Conference on {IoT} Based Control Networks and Intelligent Systems ({ICICNIS})}.\hskip 1em plus 0.5em minus 0.4em\relax IEEE, 2024, pp. 1--8.

\bibitem{li2024collm}
J.~Li, B.~Han, S.~Li, X.~Wang, and J.~Li, ``{CoLLM}: {A} {Collaborative} {LLM} {Inference} {Framework} for {Resource-Constrained} {Devices},'' in \emph{2024 {IEEE/CIC} International Conference on Communications in China ({ICCC})}.\hskip 1em plus 0.5em minus 0.4em\relax IEEE, 2024, pp. 185--190.

\bibitem{yang2024survey}
C.~Yang, Y.~Zhu, W.~Lu, Y.~Wang, Q.~Chen, C.~Gao, B.~Yan, and Y.~Chen, ``{Survey} on {Knowledge} {Distillation} for {Large} {Language} {Models}: {Methods}, {Evaluation}, and {Application},'' \emph{{ACM} Transactions on Intelligent Systems and Technology}, 2024.

\bibitem{zhu2024survey}
X.~Zhu, J.~Li, Y.~Liu, C.~Ma, and W.~Wang, ``{ASurvey} on {Model} {Compression} for {Large} {Language} {Models},'' \emph{Transactions of the Association for Computational Linguistics}, vol.~12, pp. 1556--1577, 2024.

\bibitem{liu2024minicache}
A.~Liu, J.~Liu, Z.~Pan, Y.~He, G.~Haffari, and B.~Zhuang, ``{MiniCache}: {KV} {Cache} {Compression} in {Depth} {Dimension} for {Large} {Language} {Models},'' \emph{Advances in Neural Information Processing Systems}, vol.~37, pp. 139\,997--140\,031, 2024.

\bibitem{mudvari2024splitllm}
A.~Mudvari, Y.~Jiang, and L.~Tassiulas, ``{SplitLLM}: {Collaborative} {Inference} of {LLMs} for {Model} {Placement} and {Throughput} {Optimization},'' \emph{arXiv preprint arXiv:2410.10759}, 2024.

\bibitem{chen2025llm}
X.~Chen, W.~Wu, L.~Li, and F.~Ji, ``{LLM-Empowered} {IoT} for {6G} {Networks}: {Architecture}, {Challenges}, and {Solutions},'' \emph{{IEEE} Internet of Things Magazine}, 2025.

\bibitem{he2025large}
X.~He, Y.~Jiang, X.~Xu, H.~Cui, Y.~Liu, M.~Chen, Y.~Hong, and J.~Zhang, ``{Large} {Language} {Model} {Offloading} using {Active} {Inference} in {6G} {Symbiotic} {IoT},'' \emph{{IEEE} Internet of Things Journal}, 2025.

\bibitem{cao2024multimodal}
D.~Cao, J.~Wu, and A.~K. Bashir, ``{Multimodal} {Large} {Language} {Models} {Driven} {Privacy-Preserving} {Wireless} {Semantic} {Communication} in {6G},'' in \emph{2024 {IEEE} International Conference on Communications Workshops ({ICC} Workshops)}.\hskip 1em plus 0.5em minus 0.4em\relax IEEE, 2024, pp. 171--176.

\bibitem{long20246g}
S.~Long, F.~Tang, Y.~Li, T.~Tan, Z.~Jin, M.~Zhao, and N.~Kato, ``{6G} comprehensive intelligence: {Network} operations and optimization based on {Large} {Language} {Models},'' \emph{{IEEE} Network}, 2024.

\bibitem{vaswani2017attention}
A.~Vaswani, N.~Shazeer, N.~Parmar, J.~Uszkoreit, L.~Jones, A.~N. Gomez, {\L}.~Kaiser, and I.~Polosukhin, ``{Attention} {Is} {All} {You} {Need},'' \emph{Advances in neural information processing systems}, vol.~30, 2017.

\bibitem{achiam2023gpt}
J.~Achiam, S.~Adler, S.~Agarwal, L.~Ahmad, I.~Akkaya, F.~L. Aleman, D.~Almeida, J.~Altenschmidt, S.~Altman, S.~Anadkat \emph{et~al.}, ``{Gpt-4} {Technical} {Report},'' \emph{arXiv preprint arXiv:2303.08774}, 2023.

\bibitem{ong2024efficient}
I.~Ong, ``{Efficient} {Distributed} {LLM} {Inference} with {Dynamic} {Partitioning},'' \emph{California, Berkeley, Technical Report {UCB/EECS}-2024-108, May}, 2024.

\bibitem{liu2024resource}
C.~Liu and J.~Zhao, ``{Resource} {Allocation} in {Large} {Language} {Model} {Integrated} {6G} {Vehicular} {Networks},'' in \emph{2024 {IEEE} 99th Vehicular Technology Conference ({VTC}2024-Spring)}.\hskip 1em plus 0.5em minus 0.4em\relax IEEE, 2024, pp. 1--6.

\bibitem{qian2024user}
L.~Qian and J.~Zhao, ``{User} {Association} and {Resource} {Allocation} in {Large} {Language} {Model} {Based} {Mobile} {Edge} {Computing} {System} over {6G} {Wireless} {Communications},'' in \emph{2024 {IEEE} 99th Vehicular Technology Conference ({VTC}2024-Spring)}.\hskip 1em plus 0.5em minus 0.4em\relax IEEE, 2024, pp. 1--7.

\bibitem{haider2025llm}
M.~Haider, I.~Ahmed, Z.~Hassan, K.~Hasan, and H.~V. Poor, ``{LLM-Integrated} {Digital} {Twins} for {Hierarchical} {Resource} {Allocation} in {6G} {Networks},'' \emph{arXiv preprint arXiv:2506.18293}, 2025.

\bibitem{lin2024awq}
J.~Lin, J.~Tang, H.~Tang, S.~Yang, W.-M. Chen, W.-C. Wang, G.~Xiao, X.~Dang, C.~Gan, and S.~Han, ``Awq: Activation-aware weight quantization for llm compression and acceleration,'' \emph{Proceedings of machine learning and systems}, vol.~6, pp. 87--100, 2024.

\bibitem{shen2024agile}
X.~Shen, P.~Dong, L.~Lu, Z.~Kong, Z.~Li, M.~Lin, C.~Wu, and Y.~Wang, ``Agile-quant: Activation-guided quantization for faster inference of llms on the edge,'' in \emph{Proceedings of the {AAAI} Conference on Artificial Intelligence}, vol.~38, no.~17, 2024, pp. 18\,944--18\,951.

\bibitem{topk}
S.~Zhang, G.~Cheng, W.~Wu, X.~Huang, L.~Song, and X.~Shen, ``Split fine-tuning for large language models in wireless networks,'' \emph{{IEEE} Journal of Selected Topics in Signal Processing}, pp. 1--16, 2025.

\bibitem{fwsvd}
\BIBentryALTinterwordspacing
Y.~Hsu, T.~Hua, S.~Chang, Q.~Lou, Y.~Shen, and H.~Jin, ``Language model compression with weighted low-rank factorization,'' in \emph{The Tenth International Conference on Learning Representations, {ICLR} 2022, Virtual Event, April 25-29, 2022}.\hskip 1em plus 0.5em minus 0.4em\relax OpenReview.net, 2022. [Online]. Available: \url{https://openreview.net/forum?id=uPv9Y3gmAI5}
\BIBentrySTDinterwordspacing

\bibitem{asvd}
Z.~Yuan, Y.~Shang, Y.~Song, Q.~Wu, Y.~Yan, and G.~Sun, ``Asvd: Activation-aware singular value decomposition for compressing large language models,'' \emph{arXiv preprint arXiv:2312.05821}, 2023.

\bibitem{svd-llm}
\BIBentryALTinterwordspacing
X.~Wang, Y.~Zheng, Z.~Wan, and M.~Zhang, ``{SVD-LLM:} truncation-aware singular value decomposition for large language model compression,'' in \emph{The Thirteenth International Conference on Learning Representations, {ICLR} 2025, Singapore, April 24-28, 2025}.\hskip 1em plus 0.5em minus 0.4em\relax OpenReview.net, 2025. [Online]. Available: \url{https://openreview.net/forum?id=LNYIUouhdt}
\BIBentrySTDinterwordspacing

\bibitem{lin2024duquant}
H.~Lin, H.~Xu, Y.~Wu, J.~Cui, Y.~Zhang, L.~Mou, L.~Song, Z.~Sun, and Y.~Wei, ``{Duquant}: Distributing outliers via dual transformation makes stronger quantized {llms},'' \emph{Advances in Neural Information Processing Systems}, vol.~37, pp. 87\,766--87\,800, 2024.

\bibitem{an2025systematic}
\BIBentryALTinterwordspacing
Y.~An, X.~Zhao, T.~Yu, M.~Tang, and J.~Wang, ``Systematic outliers in large language models,'' in \emph{The Thirteenth International Conference on Learning Representations, {ICLR} 2025, Singapore, April 24-28, 2025}.\hskip 1em plus 0.5em minus 0.4em\relax OpenReview.net, 2025. [Online]. Available: \url{https://openreview.net/forum?id=rLX7Vyyzus}
\BIBentrySTDinterwordspacing

\bibitem{chen2025adaptive}
Y.~Chen, R.~Li, X.~Yu, Z.~Zhao, and H.~Zhang, ``Adaptive layer splitting for wireless large language model inference in edge computing: a model-based reinforcement learning approach,'' \emph{Frontiers of Information Technology \& Electronic Engineering}, vol.~26, no.~2, pp. 278--292, 2025.

\bibitem{patel2024splitwise}
P.~Patel, E.~Choukse, C.~Zhang, A.~Shah, {\'I}.~Goiri, S.~Maleki, and R.~Bianchini, ``{Splitwise}: Efficient generative {llm} inference using phase splitting,'' in \emph{2024 {ACM/IEEE} 51st Annual International Symposium on Computer Architecture ({ISCA})}.\hskip 1em plus 0.5em minus 0.4em\relax IEEE, 2024, pp. 118--132.

\bibitem{FFT}
J.~W. Cooley and J.~W. Tukey, ``An algorithm for the machine calculation of complex fourier series,'' \emph{Mathematics of computation}, vol.~19, no.~90, pp. 297--301, 1965.

\bibitem{qin2021fcanet}
Z.~Qin, P.~Zhang, F.~Wu, and X.~Li, ``{Fcanet}: Frequency channel attention networks,'' in \emph{Proceedings of the {IEEE/CVF} international conference on computer vision}, 2021, pp. 783--792.

\bibitem{liu2024ffsplit}
Z.~Liu, Q.~Song, Q.~C. Xiao, S.~K. Selvaraj, R.~Mazumder, A.~Gupta, and X.~Hu, ``{Ffsplit}: Split feed-forward network for optimizing accuracy-efficiency trade-off in language model inference,'' \emph{arXiv preprint arXiv:2401.04044}, 2024.

\bibitem{li2018constrained}
C.~Li and C.~Shi, ``Constrained optimization based low-rank approximation of deep neural networks,'' in \emph{Proceedings of the European Conference on Computer Vision ({ECCV})}, 2018, pp. 732--747.

\bibitem{liu2018frequency}
Z.~Liu, J.~Xu, X.~Peng, and R.~Xiong, ``Frequency-domain dynamic pruning for convolutional neural networks,'' \emph{Advances in neural information processing systems}, vol.~31, 2018.

\bibitem{liu2024spinquant}
\BIBentryALTinterwordspacing
Z.~Liu, C.~Zhao, I.~Fedorov, B.~Soran, D.~Choudhary, R.~Krishnamoorthi, V.~Chandra, Y.~Tian, and T.~Blankevoort, ``Spinquant: {LLM} quantization with learned rotations,'' in \emph{The Thirteenth International Conference on Learning Representations, {ICLR} 2025, Singapore, April 24-28, 2025}.\hskip 1em plus 0.5em minus 0.4em\relax OpenReview.net, 2025. [Online]. Available: \url{https://openreview.net/forum?id=ogO6DGE6FZ}
\BIBentrySTDinterwordspacing

\bibitem{zhao2024atom}
Y.~Zhao, C.-Y. Lin, K.~Zhu, Z.~Ye, L.~Chen, S.~Zheng, L.~Ceze, A.~Krishnamurthy, T.~Chen, and B.~Kasikci, ``{Atom}: Low-bit quantization for efficient and accurate {llm} serving,'' \emph{Proceedings of Machine Learning and Systems}, vol.~6, pp. 196--209, 2024.

\bibitem{wang2025exploring}
Y.~Wang, D.~Dai, Z.~Yang, J.~Ma, and Z.~Sui, ``Exploring activation patterns of parameters in language models,'' in \emph{Proceedings of the {AAAI} Conference on Artificial Intelligence}, vol.~39, no.~24, 2025, pp. 25\,416--25\,424.

\bibitem{li2024adaptive}
\BIBentryALTinterwordspacing
W.~Li, L.~Li, M.~G. Lee, and S.~Sun, ``Adaptive layer sparsity for large language models via activation correlation assessment,'' in \emph{The Thirty-eighth Annual Conference on Neural Information Processing Systems}, 2024. [Online]. Available: \url{https://openreview.net/forum?id=Jup0qZxH7U}
\BIBentrySTDinterwordspacing

\bibitem{DBLP:journals/corr/abs-2503-06518}
\BIBentryALTinterwordspacing
F.~Zhang, Y.~Liu, W.~Li, J.~Lv, X.~Wang, and Q.~Bai, ``Towards superior quantization accuracy: {A} layer-sensitive approach,'' \emph{CoRR}, vol. abs/2503.06518, 2025. [Online]. Available: \url{https://doi.org/10.48550/arXiv.2503.06518}
\BIBentrySTDinterwordspacing

\bibitem{llama3modelcard}
\BIBentryALTinterwordspacing
{AI@Meta}, ``{Llama 3 Model Card},'' 2024. [Online]. Available: \url{https://github.com/meta-llama/llama3/blob/main/MODEL_CARD.md}
\BIBentrySTDinterwordspacing

\bibitem{qwen2.5}
\BIBentryALTinterwordspacing
{Qwen Team}, ``{Qwen2.5}: A party of foundation models,'' September 2024. [Online]. Available: \url{https://qwenlm.github.io/blog/qwen2.5/}
\BIBentrySTDinterwordspacing

\bibitem{OpenBookQA2018}
T.~Mihaylov, P.~Clark, T.~Khot, and A.~Sabharwal, ``{Can} a {Suit} of {Armor} {Conduct} {Electricity}? {A} {New} {Dataset} for {Open} {Book} {Question} {Answering},'' in \emph{{EMNLP}}, 2018.

\bibitem{allenai:arc}
P.~Clark, I.~Cowhey, O.~Etzioni, T.~Khot, A.~Sabharwal, C.~Schoenick, and O.~Tafjord, ``Think you have solved question answering? {Try} {ARC}, the {AI2} reasoning challenge,'' \emph{arXiv:1803.05457v1}, 2018.

\bibitem{Bisk2020}
Y.~Bisk, R.~Zellers, R.~L. Bras, J.~Gao, and Y.~Choi, ``{PIQA}: Reasoning about physical commonsense in natural language,'' in \emph{Thirty-Fourth {AAAI} Conference on Artificial Intelligence}, 2020.

\bibitem{sap2019socialiqa}
M.~Sap, H.~Rashkin, D.~Chen, R.~LeBras, and Y.~Choi, ``{SocialIQA}: Commonsense reasoning about social interactions,'' 2019.

\bibitem{ai2:winogrande}
``{WinoGrande}: An adversarial {Winograd} schema challenge at scale,'' 2019.

\bibitem{talmor-etal-2019-commonsenseqa}
\BIBentryALTinterwordspacing
A.~Talmor, J.~Herzig, N.~Lourie, and J.~Berant, ``{C}ommonsense{QA}: A question answering challenge targeting commonsense knowledge,'' in \emph{Proceedings of the 2019 Conference of the North {A}merican Chapter of the Association for Computational Linguistics: Human Language Technologies, Volume 1 (Long and Short Papers)}.\hskip 1em plus 0.5em minus 0.4em\relax Minneapolis, Minnesota: Association for Computational Linguistics, Jun. 2019, pp. 4149--4158. [Online]. Available: \url{https://aclanthology.org/N19-1421}
\BIBentrySTDinterwordspacing

\bibitem{allenai:qasc}
\BIBentryALTinterwordspacing
T.~Khot, P.~Clark, M.~Guerquin, P.~Jansen, and A.~Sabharwal, ``{QASC:} {A} dataset for question answering via sentence composition,'' in \emph{The Thirty-Fourth {AAAI} Conference on Artificial Intelligence, {AAAI} 2020, The Thirty-Second Innovative Applications of Artificial Intelligence Conference, {IAAI} 2020, The Tenth {AAAI} Symposium on Educational Advances in Artificial Intelligence, {EAAI} 2020, New York, NY, USA, February 7-12, 2020}.\hskip 1em plus 0.5em minus 0.4em\relax {AAAI} Press, 2020, pp. 8082--8090. [Online]. Available: \url{https://doi.org/10.1609/aaai.v34i05.6319}
\BIBentrySTDinterwordspacing

\bibitem{liu2020logiqa}
\BIBentryALTinterwordspacing
J.~Liu, L.~Cui, H.~Liu, D.~Huang, Y.~Wang, and Y.~Zhang, ``Logiqa: {A} challenge dataset for machine reading comprehension with logical reasoning,'' in \emph{Proceedings of the Twenty-Ninth International Joint Conference on Artificial Intelligence, {IJCAI} 2020}, C.~Bessiere, Ed.\hskip 1em plus 0.5em minus 0.4em\relax ijcai.org, 2020, pp. 3622--3628. [Online]. Available: \url{https://doi.org/10.24963/ijcai.2020/501}
\BIBentrySTDinterwordspacing

\bibitem{huang-etal-2019-cosmos}
\BIBentryALTinterwordspacing
L.~Huang, R.~Le~Bras, C.~Bhagavatula, and Y.~Choi, ``{Cosmos} {QA}: Machine reading comprehension with contextual commonsense reasoning,'' in \emph{Proceedings of the 2019 Conference on Empirical Methods in Natural Language Processing and the 9th International Joint Conference on Natural Language Processing ({EMNLP-IJCNLP})}.\hskip 1em plus 0.5em minus 0.4em\relax Hong Kong, China: Association for Computational Linguistics, Nov. 2019, pp. 2391--2401. [Online]. Available: \url{https://www.aclweb.org/anthology/D19-1243}
\BIBentrySTDinterwordspacing

\bibitem{Zhuang2024Medical}
Z.~Zhuang, Z.~Zhuang, and T.~Wang, ``Medical image encryption algorithm based on a new five-dimensional multi-band multi-wing chaotic system and {QR} decomposition,'' \emph{Scientific Reports}, vol.~14, no.~1, p. 402, 2024.

\end{thebibliography}

\vspace{3ex}
\section*{Biographies}

\begin{IEEEbiography}[{\includegraphics[width=1in,height=1.25in,clip,keepaspectratio]{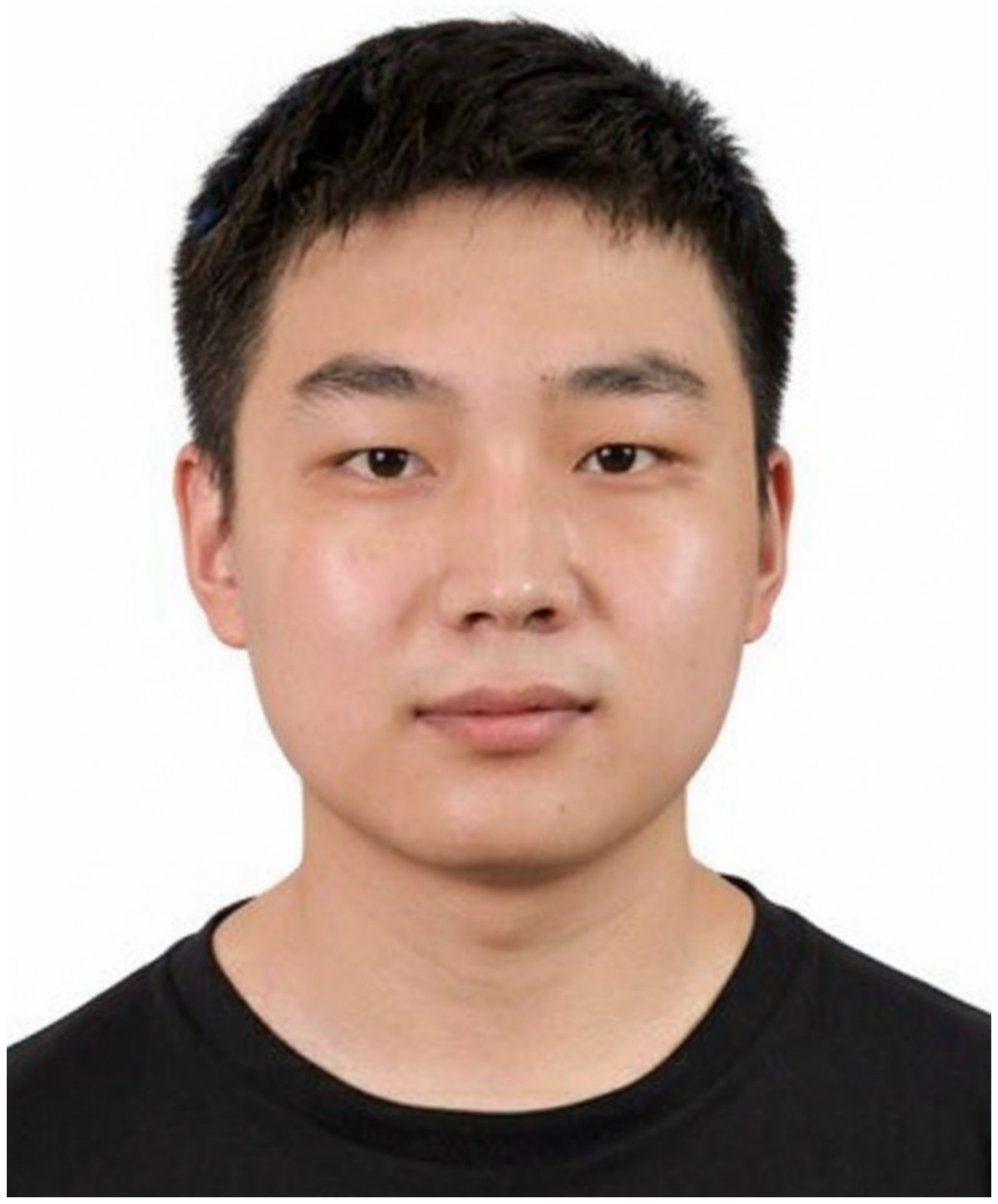}}]{JIAN MA}
(mj@bupt.edu.cn) is currently pursuing the Ph.D. degree with the School of Cyberspace Security at the Beijing University of Posts and Telecommunications (BUPT). His research interests include edge intelligence, split learning, and artificial intelligence.
\end{IEEEbiography}

\vspace{5\baselineskip} 


\begin{IEEEbiography}[{\includegraphics[width=1in,height=1.25in,clip,keepaspectratio]{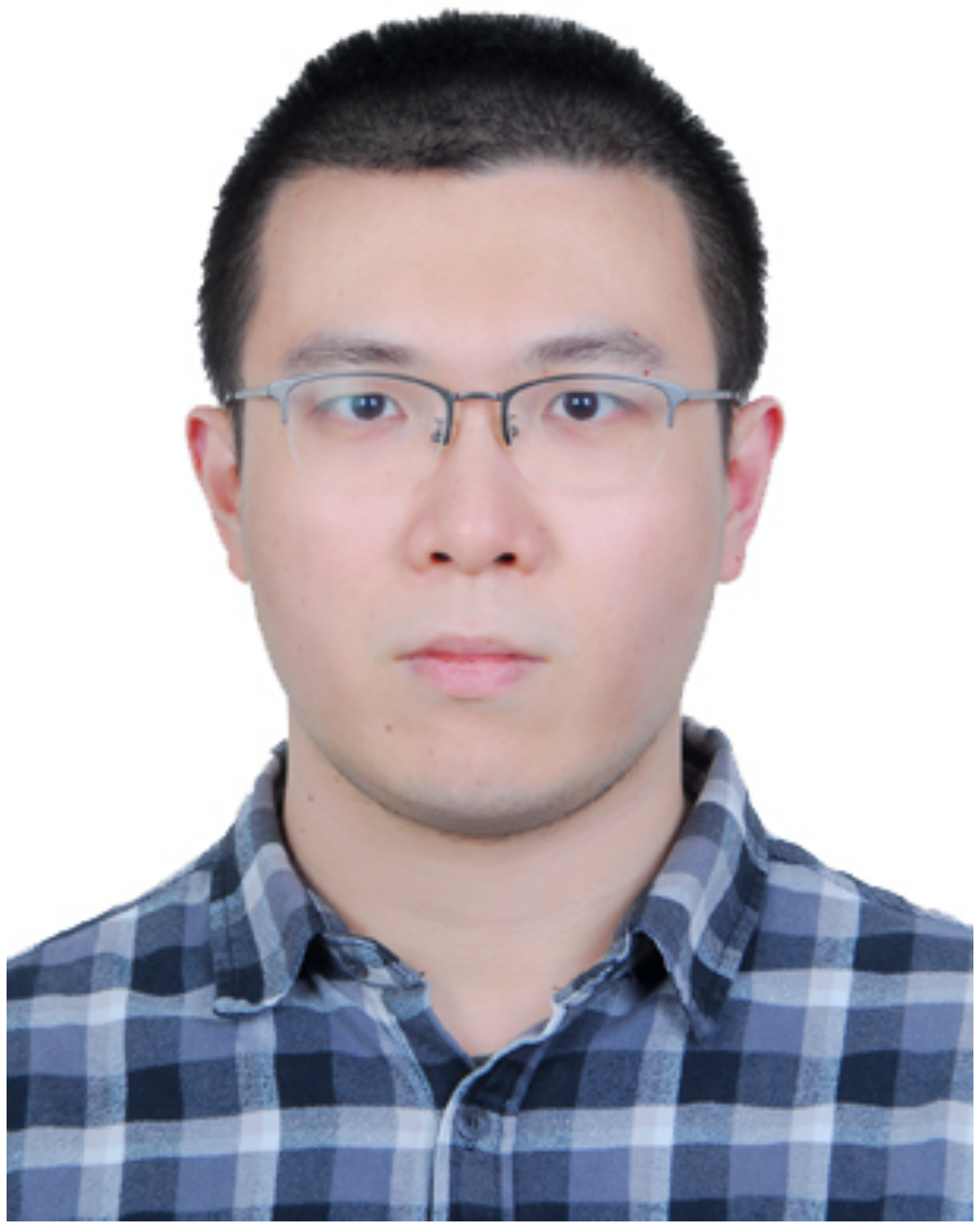}}]{XINCHEN LYU} (lvxinchen@bupt.edu.cn) received the B.E. degree from BUPT in 2014, and the dual Ph.D. degrees from BUPT and the University of Technology Sydney in 2019. He is currently an Associate Professor with the National Engineering Research Center of Mobile Network Technologies, BUPT. His research interests include resource management and security of edge intelligence and its applications in future wireless networks.
\end{IEEEbiography}

\vspace{-8ex} 

\begin{IEEEbiography}[{\includegraphics[width=1in,height=1.25in,clip,keepaspectratio]{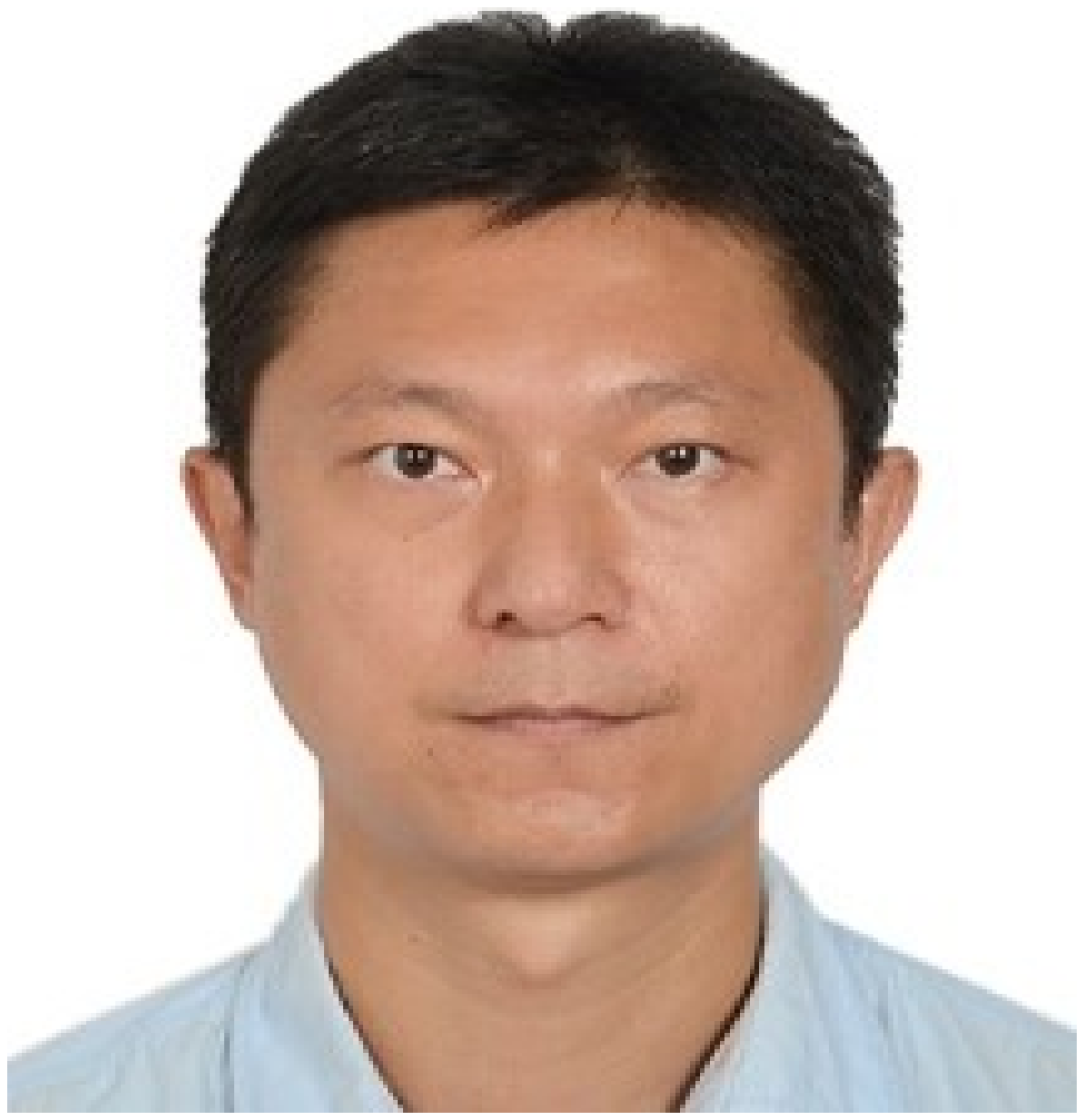}}]{JUN JIANG}
(jiangj@pcl.ac.cn) received the Ph.D. degrees from the Harbin Institute of Technology (HIT), Harbin, China, in 2009. He is currently a Senior Engineer with the Pengcheng Laboratory (PCL), Shenzhen, China. His current research interests include 3-D computer vision, SLAM, and deep learning.
\end{IEEEbiography}

\vspace{-8ex} 

\begin{IEEEbiography}[{\includegraphics[width=1in,height=1.25in,clip,keepaspectratio]{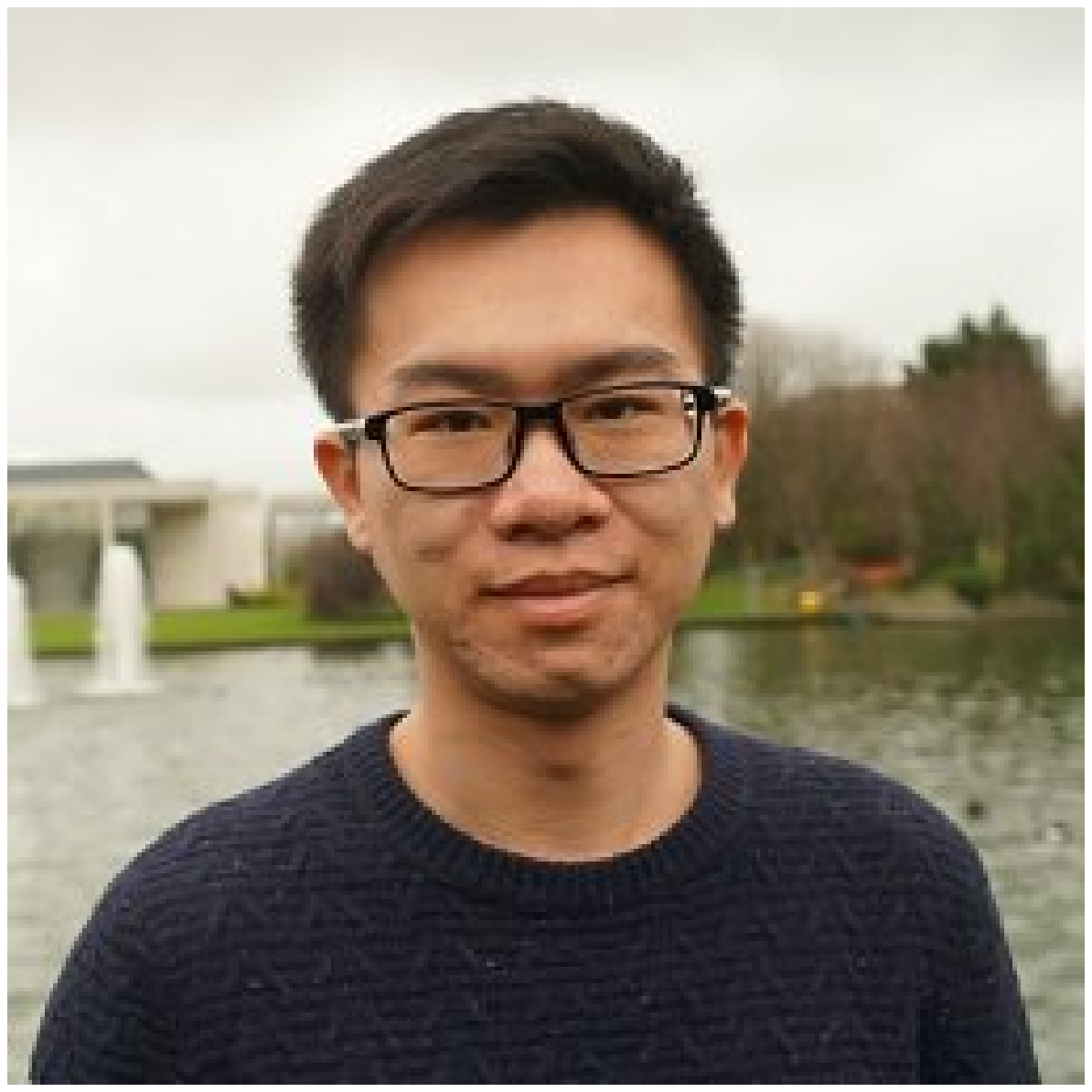}}]{LONGHAO ZOU}
(zoulh@pcl.ac.cn) received the BEng and PhD degrees from Beijing University of Posts and Telecommunications, Beijing, China, and Dublin City University (DCU), Dublin, Ireland, in 2011 and 2016, respectively. He was a postdoctoral researcher with the European Union's Horizon 2020 NEWTON Project at DCU, Dublin Ireland. Now he is an associate researcher with Peng Cheng Laboratory, Shenzhen, China, and also with Southern University of Science and Technology, Shenzhen, China. His research interests include mobile and wireless communications, holographic communication, multi-sensorial interaction, resource allocation, and user quality of experience.
\end{IEEEbiography}

\vspace{-8ex} 

\begin{IEEEbiography}[{\includegraphics[width=1in,height=1.25in,clip,keepaspectratio]{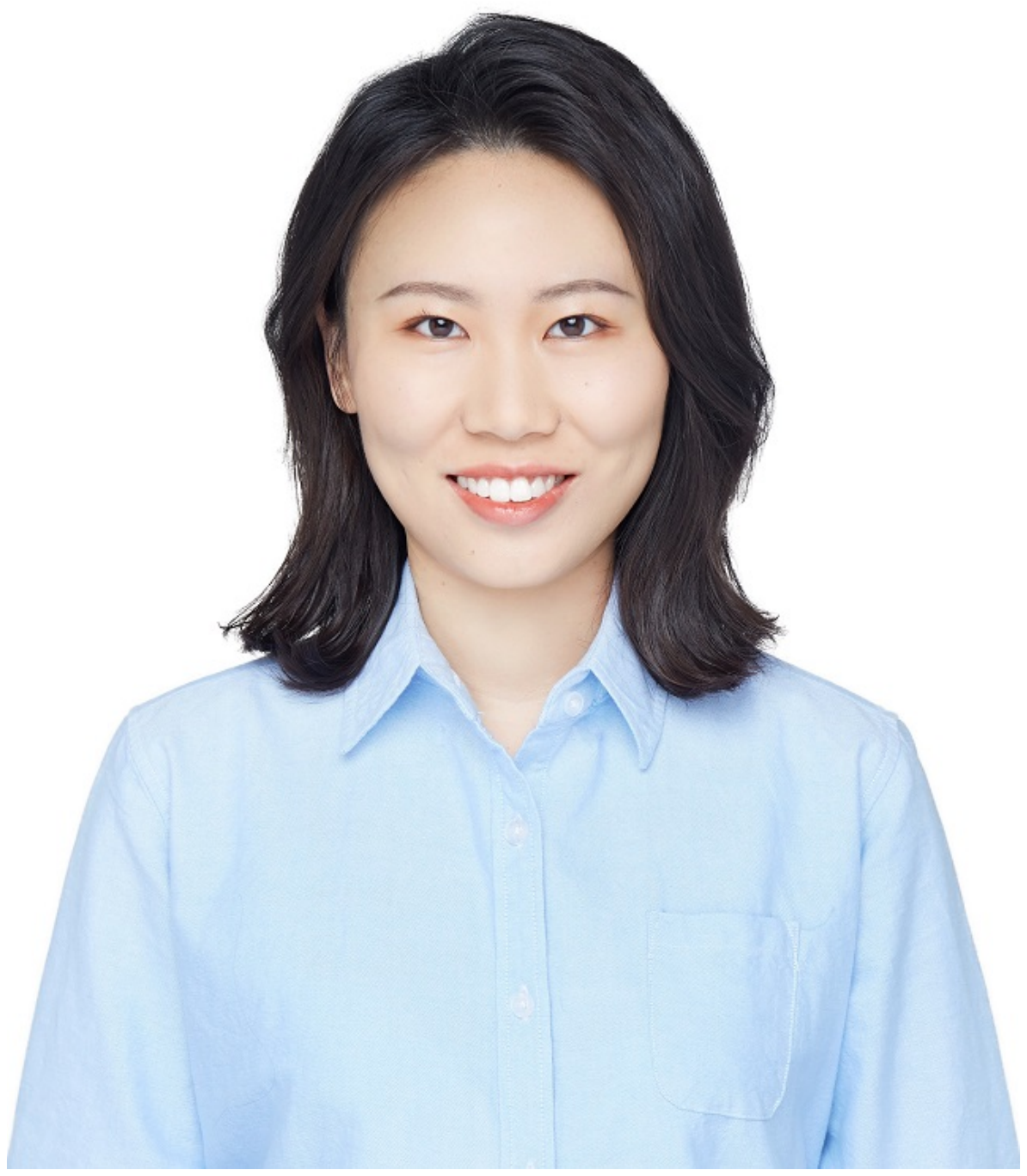}}]{CHENSHAN REN}
(renchenshan06@163.com) received the B.E. degree from Zhengzhou University, Henan, China, in 2013, the first Ph.D. degree from the Beijing University of Posts and Telecommunications, and the second Ph.D. degree from the University of Technology Sydney in 2019. She is currently a Lecturer with the Minzu University of China. Her research interests include fog computing, software-defined networking, and radio resource management.
\end{IEEEbiography}

\vspace{-8ex} 

\begin{IEEEbiography}[{\includegraphics[width=1in,height=1.25in,clip,keepaspectratio]{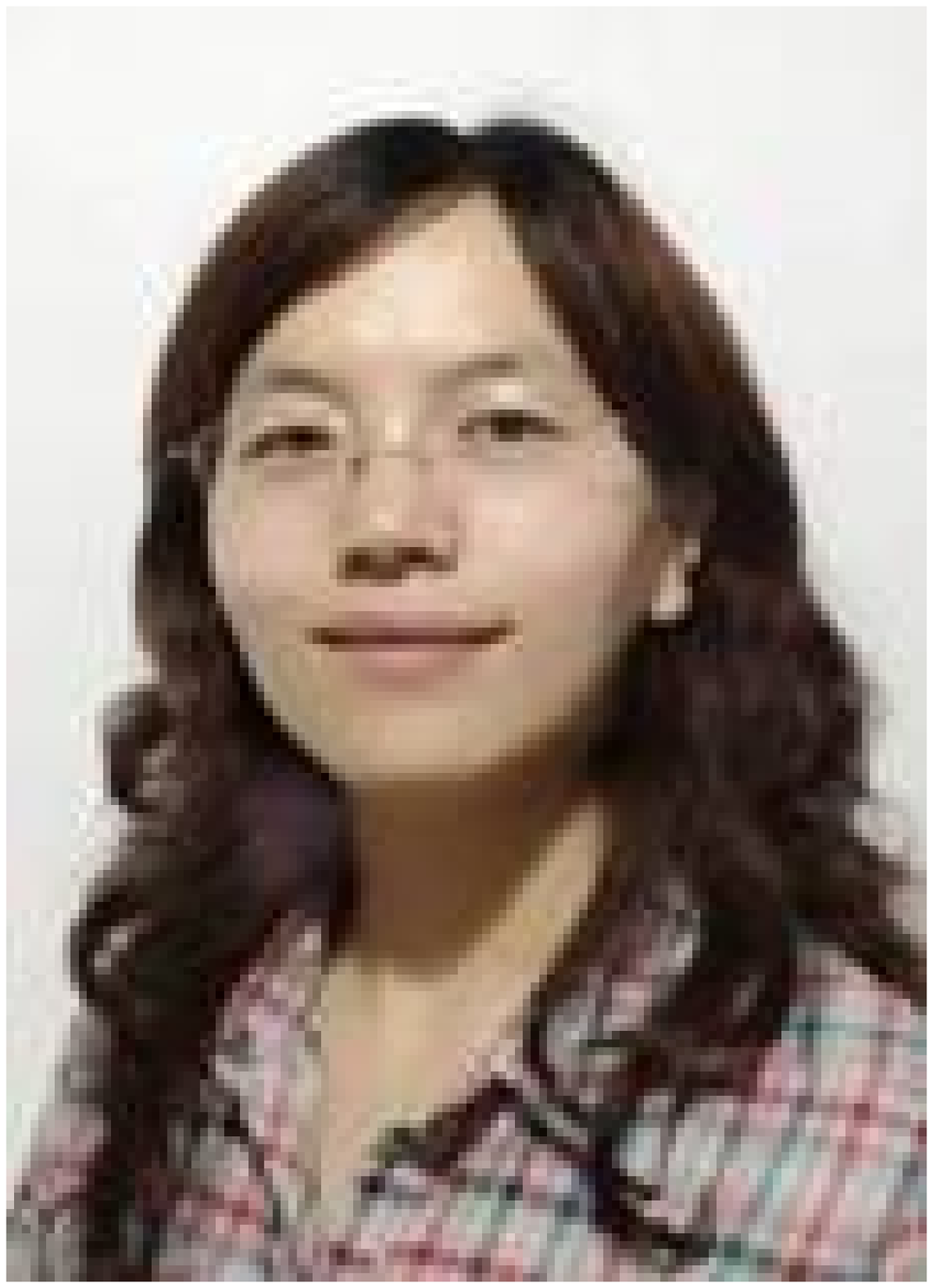}}]{QIMEI CUI}
(cuiqimei@bupt.edu.cn) received the B.E. and M.S. degrees from Hunan University in 2000 and 2003, respectively, and the Ph.D. degree from the Beijing University of Posts and Telecommunications (BUPT) in 2006. She is currently a Full Professor with the School of Information and Communication Engineering, BUPT. Her research interests include energy-efficient transmission theory and networking technology for wireless and green communications.
\end{IEEEbiography}

\vspace{-8ex} 

\begin{IEEEbiography}[{\includegraphics[width=1in,height=1.25in,clip,keepaspectratio]{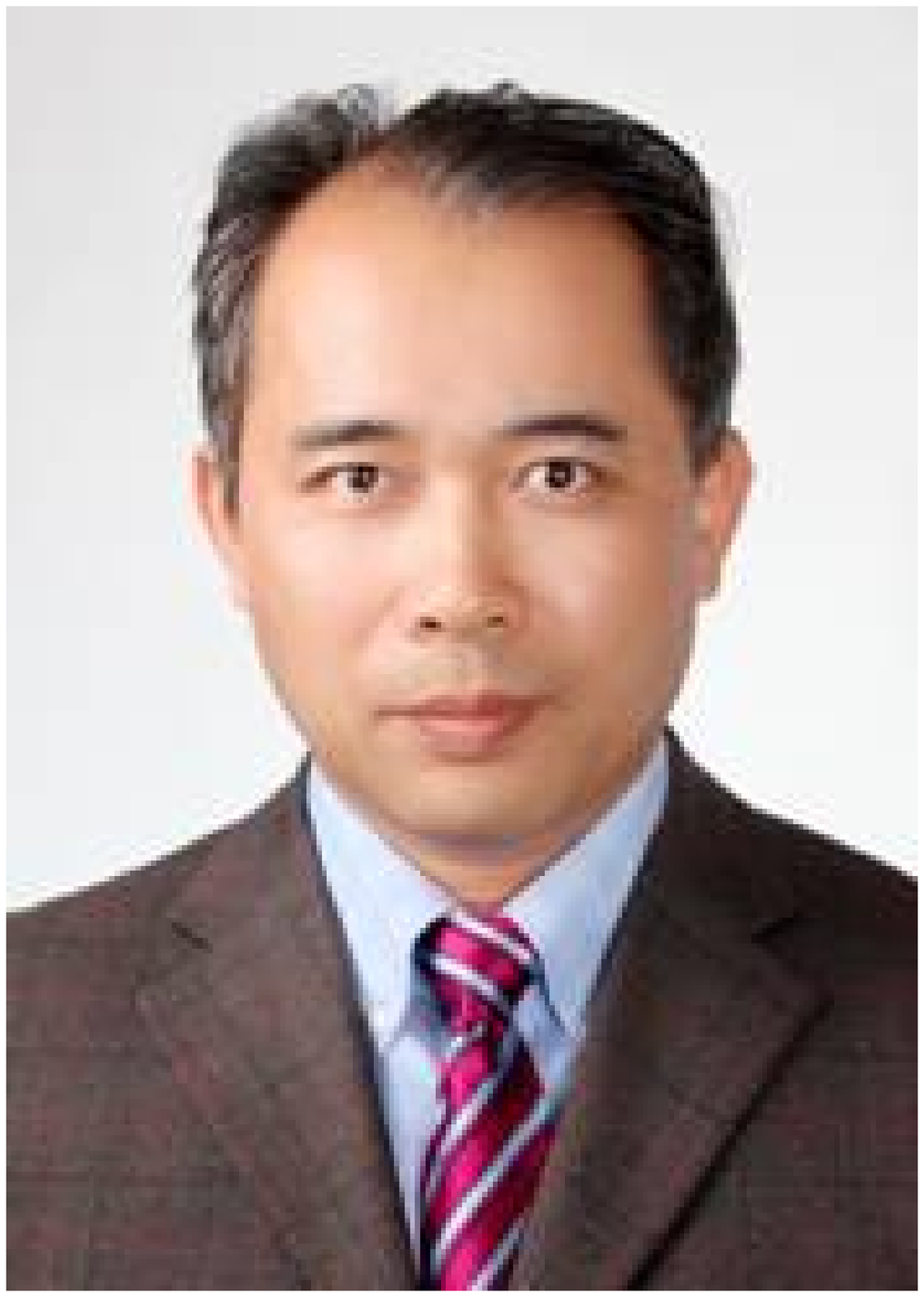}}]{XIAOFENG TAO}
(taoxf@bupt.edu.cn) received the B.S. degree in electrical engineering from Xi'an Jiaotong University, Xi'an, China, in 1993, and the M.S. and Ph.D. degrees in telecommunication engineering from the Beijing University of Posts and Telecommunications (BUPT), Beijing, China, in 1999 and 2002, respectively. He is currently a Full Professor with BUPT, a fellow of the Institution of Engineering and Technology, and the Chair of the IEEE ComSoc Beijing Chapter. He is also with the Pengcheng Laboratory (PCL), Shenzhen, China.
\end{IEEEbiography}

\end{document}